\documentclass[sigconf,nonacm,letterpaper]{acmart}

\AtBeginDocument{%
  \providecommand\BibTeX{{%
    \normalfont B\kern-0.5em{\scshape i\kern-0.25em b}\kern-0.8em\TeX}}}

\usepackage{multirow}
\usepackage{graphicx}
\usepackage{svg}
\usepackage{caption}
\usepackage{xcolor}
\usepackage{subcaption}
\usepackage{pifont}
\usepackage{enumitem}
\usepackage{siunitx}
\usepackage{tcolorbox}

\newcommand{\cmark}{\ding{51}}%
\newcommand{\xmark}{\ding{55}}%

\begin{document}

\title{Zen: LSTM-based generation of individual  spatiotemporal cellular traffic with interactions}

\author{Anne Josiane Kouam}
\affiliation{%
  \institution{Inria}
  \country{France}
  }

\author{Aline Carneiro Viana}
\affiliation{%
  \institution{Inria}
  \country{France}
}

\author{Alain Tchana}
\affiliation{%
 \institution{Grenoble INP}
 \country{France}
 }

\renewcommand{\shortauthors}{Kouam et al.}

\begin{abstract}
Domain-wide recognized by their high value in human presence and activity studies, cellular network datasets (i.e., Charging Data Records, named CdRs), however, present accessibility, usability, and privacy issues, restricting their exploitation and research reproducibility. 
This paper tackles such challenges by modeling Cdrs that fulfill real-world data attributes. 
Our designed framework, named \textit{Zen} follows a four-fold methodology related to (i) the LTSM-based modeling of users' traffic behavior, (ii) the realistic and flexible emulation of spatiotemporal mobility behavior, (iii) the structure of lifelike cellular network infrastructure and social interactions, and (iv) the combination of the three previous modules into realistic Cdrs traces with an individual basis, realistically.
Results show that \textit{Zen}'s first and third models accurately capture individual and global distributions of a fully anonymized real-world Cdrs dataset, while the second model is consistent with the literature's revealed features in human mobility.
Finally, we validate \textit{Zen} Cdrs ability of reproducing daily cellular behaviors of the urban population and its usefulness in practical networking applications 
such as dynamic population tracing, Radio Access Network's power savings, and anomaly detection as compared to real-world CdRs.


\end{abstract}

\ccsdesc[500]{Data and Communication Traffic~Charging Data Records}
\ccsdesc[300]{Cellular Traffic~Mobility and Network events}
\ccsdesc[500]{Modeling~LSTM}

\keywords{Human mobility modeling, Data and Communication traffic modeling, Recurrent Neural Networks.}

\maketitle

\section{Introduction} 
\label{sec:intro}


\begin{figure}
    \centering
    \includegraphics[width=0.9\linewidth]{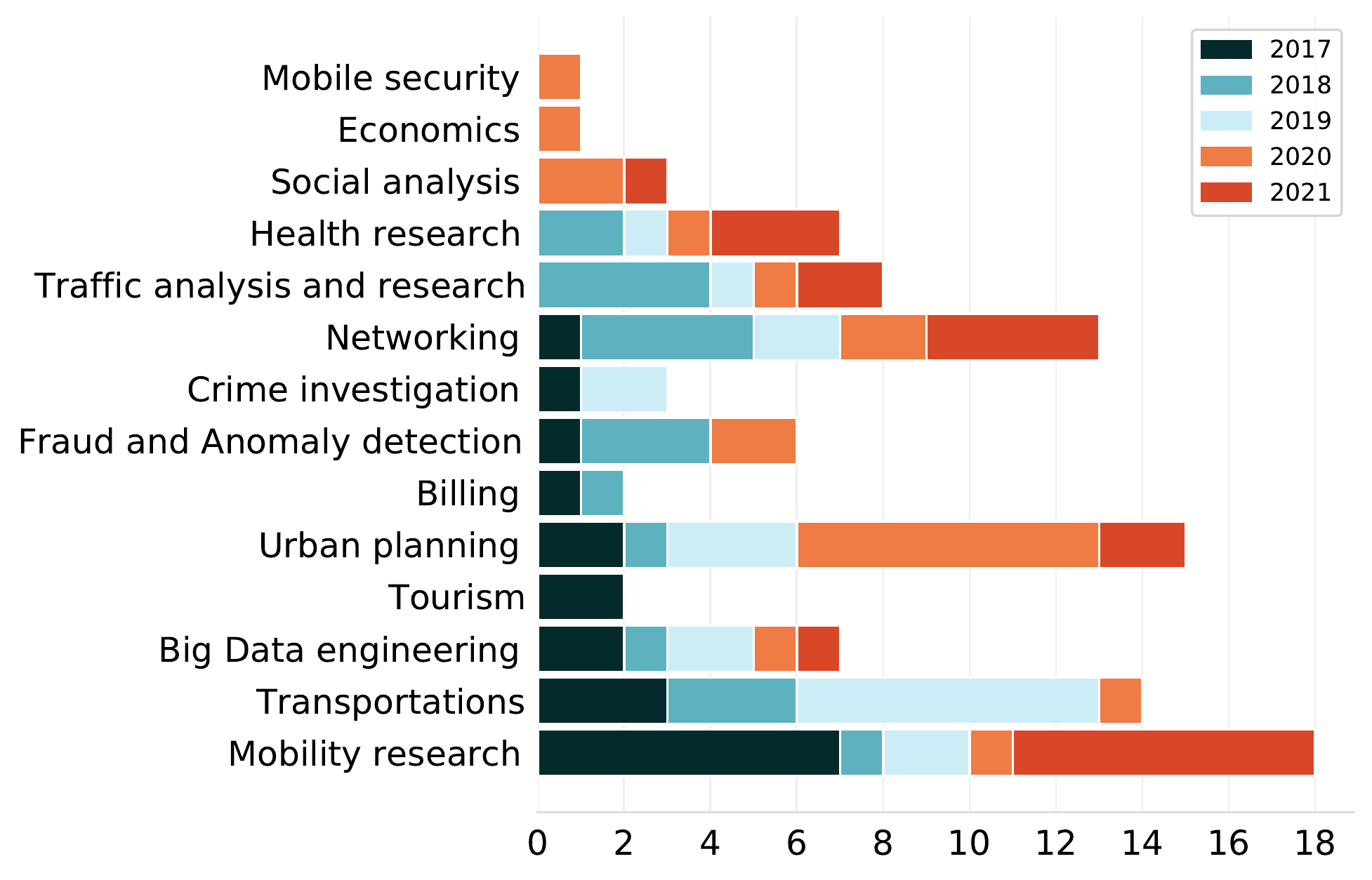}
    \caption{Distribution by domain of the last 5-year most relevant publications using CdRs.}
    \label{fig:CdRs_usage}
\end{figure}


Charging Data Records are acknowledged as a common tool for studying human mobility, infrastructure usage, and traffic behavior~\cite{Jiang:2013}. 
We name such datasets as CdRs to distinguish them from the standard Call Detail Records (CDRs), describing only call and SMS cellular communication information.
CdRs describe time-stamped and geo-referenced event types (i.e., data, calls, SMS) generated by each mobile device interacting with operator networks (cf. Table~\ref{tab:CdR}). They comprise city-, region-, or country-wide areas and usually cover long periods (months or years); no other technology currently provides an equivalent per-device precise scope. 
As a result, CdRs are exploited in different research domains and industries, such as sociology~\cite{Rhoads:2020}, epidemiology~\cite{Chang:2021}, transportation~\cite{Qin:2017}, and networking~\cite{Ozturk:2021}.
For a quantitative appreciation of such CdRs' worth recognition, Fig.~\ref{fig:CdRs_usage} identifies 14 different research domains leveraging CdRs among 100 most relevant works (sorted by Google Scholar) selected from 1022 last 5-year publications. This clearly shows a great diversity of domains in this sample only ($\sim10\%$).


Yet, the exploitation of real-world CdRs for research faces many limitations. 
First, \textit{accessibility}: CdRs datasets are not publicly available, imposing strict mobile operators' agreements.
Second, \textit{usability}: CdRs are usually available in an aggregated form (i.e., grouped mobility flows and coarse spatiotemporal information), limiting related analyses' preciseness.
Third, \textit{privacy}: even anonymized, non-aggregated CdRs describe sensitive information of users' habits, which hardens their shareability~\cite{Montjoye:2013}.
This paper \textit{addresses such limitations by enabling the autonomous generation of realistic and privacy-compliant CdRs by scientific community, thus providing new avenues for research advances.}

Moreover, generated CdRs should conform to essential attributes, namely, \textit{completeness, realisticness, fine-grained description, and privacy. } 
Unfortunately, those attributes make the generation of realistic CdRs challenging and complex. In particular, achieving completeness requires (i) either real-world complete CdRs datasets (hard to obtain) describing mobility, traffic, and
pairwise users communications or (ii) to cope with the difficulty in modeling the intrinsic correlations between information describing users' behaviors in space, time, and social communication.
Achieving \textit{realisticness} implies considering real-world cellular network complexities (architecture and topology) at all levels of the generation process. The \textit{fine-grained description} achievement is impeded by the heterogeneity of users' behaviors, especially in cellular traffic. 
Finally, generated traces should be \textit{privacy-compliant} to avoid backtracking real users' identities, most often done through their mobility information. 

To the best of our knowledge, \textit{this is the first work in literature producing 
realistic Charging Data Records (CdRs) that fulfill the above-mentioned attributes}. 
Our designed framework, named \textit{Zen} employs a four-fold methodology:

\vspace{0.05cm}
\textbf{(1)} Leveraging on a real-world fully anonymized CdRs describing users' traffic behavior (i.e, events information on its type~-- data, call, and SMS~--, duration, pairwise information, etc), \textit{we propose the first literature modeling 
that captures long-range and inter-CdRs specificity} correlations while addressing the population heterogeneity (\S \ref{sec:traffic}). Our model captures population diverseness 
in the reproduction of individual traffic behaviors. We use three separate \textit{Long-Short-Term Memory neural networks} (LSTM) to model event types generation (i.e., \textit{what}), the inter-event duration (i.e., \textit{when}), the social interactions (i.e., \textit{whom}), and leverage statistical analysis to model CdRs metrics such as calls duration (\textit{i.e., how}). Overall, \textit{Zen} traffic modeling presents significant high performance values, providing for 80\% of users (i) more than 95\% (for event-type) and 75\% (for inter-event) of modeling accuracy, and
(ii) less than 6.68\% (for inter-event) and 12.5\% (for social) of Mean Absolute Error's maximum values.


\vspace{0.05cm}
\textbf{(2)} 
Mobility behaviors of individuals (\S \ref{sec:mobility}) are emulated according to the infrastructure of a real-world metropolitan city (here, the Helsinki EU city), and resulting trajectories are coupled with the corresponding cell towers distribution of existing operator networks in the same city \cite{OpenCellID}.
Here, we leverage city planning, transportation information \cite{HelsinkiReport} as well literature investigations on laws dictating human mobility \cite{Amichi:2020, Marta:2008, Eduardo_Mob:2016}.
Such real-world information and realistic human mobility modeling are then incorporated in the literature \textit{Working Day Mobility} (WDM) model~\cite{WDM}~-- extensively enhancing it 
~-- to emulate urban daily-life mobility behaviors of individuals.
Moreover, we rely on the ONE simulator~\cite{ONE} to bring flexibility to our model regarding population size, duration, and covered area.

\vspace{0.05cm}
\textbf{(3)} We then design a separate module (\S \ref{sec:social_ties}) to realistically reproduce on top of generated mobility traces, a cellular network organization with multiple operators 
and build social ties between users. This enables the first-of-a-kind flexibility to produce CdRs of numerous operators at the same period. 

\vspace{0.05cm}
\textbf{(4)} We combine all the previous models to generate complete CdRs describing individual mobility, traffic, pairwise communications following real traffic behavior (\S \ref{sec:zen-merger}). 


Note that the disjoint behaviors modeling of realistic emulated mobility and of real-world traffic hides real individuals' spatiotemporal daily-life habits in routine and leisure times (e.g., home/work, nightlife, etc.), bringing the privacy-preserving capability to the produced CdRs.

\section{Zen Overview}
\label{sec:overview}

In the following, we provide an overview of \textit{Zen} architecture and describe the different real-world datasets we leverage.

\subsection{Architecture}
\label{subsec:zen_description}

\begin{figure}
    \centering
    \includegraphics[width=1.05\linewidth]{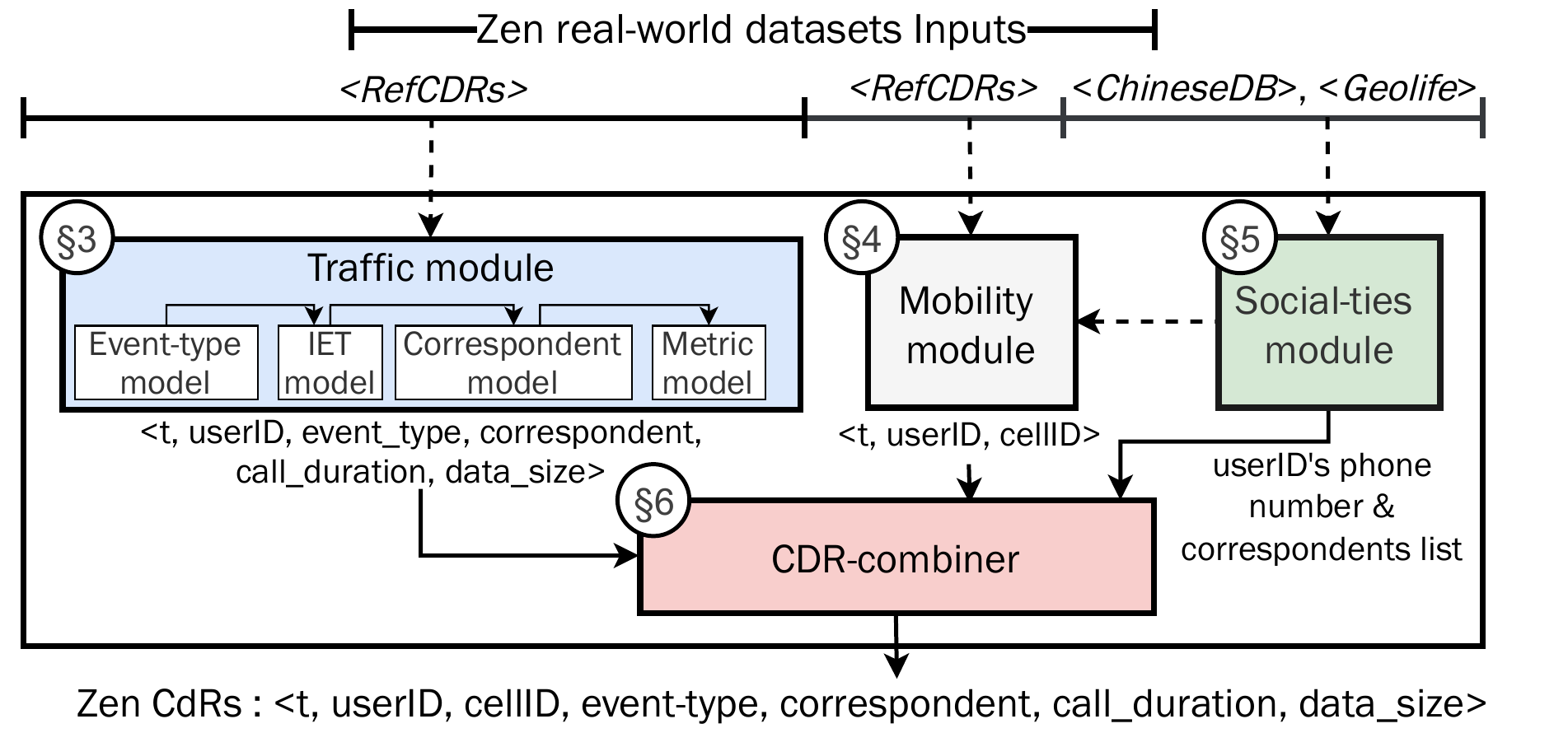}
    \caption{\textit{Zen} architecture.} 
    \label{fig:zen_overview}
\end{figure}

According to input parameters, we generate realistic CdRs (cf. Table \ref{tab:CdR}) through four phases, each implemented in a module of the \textit{Zen} architecture (cf. Fig. \ref{fig:zen_overview}). \textit{Zen} architecture consists of (1) a \textit{traffic module}, (2) a \textit{mobility module}; (3) a \textit{social-ties module}, and (4) a \textit{CdR-combiner}, or merger module. 

The \textit{traffic module} (\S \ref{sec:traffic}) leverages \textit{Long-Short-Term Memory neural networks (LSTM)} jointly with statistical analysis to model users' traffic behavior from real-world CdRs. It provides answers to \textit{what}, \textit{when}, with \textit{whom}, and \textit{how} to generate events.
The \textit{mobility module} (\S \ref{sec:mobility}) (i) emulates users temporal displacements on a real-world geographical map over a selected period, and (ii) associates corresponding users positions with a real-world cellular topology.
This dataset feeds the \textit{social-ties module} (\S \ref{sec:social_ties}) that builds the network social structure on top of which users' communication interactions occur by building users' phonebooks, i.e., list of phone numbers a user is likely to contact.
Finally, the \textit{CdR-combiner} module (\S \ref{sec:zen-merger}) combines the previous modules' outputs to generate realistic CdRs per network operator over a specified duration and particular urban area.


\subsection{Real-world reference datasets}
\label{subsec:datasets}

\textit{Zen} models real-world datasets to produce realistic outputs. In particular, as depicted on top of Fig. \ref{fig:zen_overview}, \textit{Zen} uses three real-world reference datasets described in what follows. 

\textit{RefCdRs} are used by both the \textit{traffic} and the \textit{social-ties} modules. \textit{RefCdRs} refer to a fully-anonymized CdRs dataset collected by a major mobile network operator. They describe 1-month (\textit{from 2018-06-01 to 2018-06-30}) per-user traffic resulting in about 3 million timestamped events generated by 186,738 distinct phone numbers, where about 17,000 are from the \textit{RefCdRs}' operator. 
\textit{RefCdRs} are incomplete; they lack mobility features and incoming-SMS traffic type (i.e., only have outgoing SMS). Still, there is no information on the size of sessions in the data traffic type. 
\textit{RefCdRs} provide each user's operator network code. We leverage this information to identify the list of operators appearing in the datasets. \\
On the other hand, the \textit{mobility module} leverages the \textit{ChineseDB}~\cite{Amichi:2020} and \textit{Geolife}~\cite{Geolife1} datasets, by extracting statistics describing real-life mobility behavior of users. \textit{ChineseDB} (non-public and fully anonymized mobility CdRs) contains trajectories of 642K users during two weeks.  In particular, we did not have access to \textit{ChineseDB} but only to related statistics available in \cite{Amichi:2020}. \textit{Geolife} (public and anonymized GPS dataset) contains trajectories of 182 users during 64 months. 

\subsection{\textit{Zen} CdRs attributes}
\label{subsec:datasets}
We present hereafter the positioning of \textit{Zen} generated CdRs with respect to our goals:

\vspace{0.06cm}
\noindent\textbf{\textit{Completeness:}}
Complete CdRs comprise mobility and traffic features and should, thus, include, in addition to user positions (i.e., network cell Ids), all event types, namely data, call, and SMS.
Here, the limited access to complete real-world CdRs hardens the modeling and reproduction of complete CdRs. 
\textit{Zen} circumvents this limitation and provides complete CdRs by jointly modeling separate CdRs features to capture the implicit correlations between them : e.g., the choice of \textit{whom} to communicate with is generally time (\textit{when}) and event (\textit{what}) dependent.
Therefore, Table \ref{tab:CdR_completeness} shows \textit{Zen} yields complete CdRs compared to most state-of-the-art contributions which instead provide only one CdRs feature, either mobility or an event type.

\vspace{0.06cm}
\noindent\textbf{\textit{Realisticness:}}
\textit{Zen} modeling integrates a real telecom network topology (\textit{inducing users' cell-tower locations}) and organization in multi-operators, as a requirement to produce realistic CdRs. Considering the latter, Zen CdRs conveys inter-operator interaction patterns, valuable for telecom fraud investigation for instance~\cite{Kouam:2021}.

\begin{figure}
\includegraphics[scale=0.24]{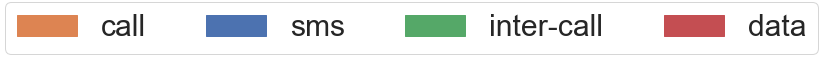}
\includegraphics[scale=0.15]{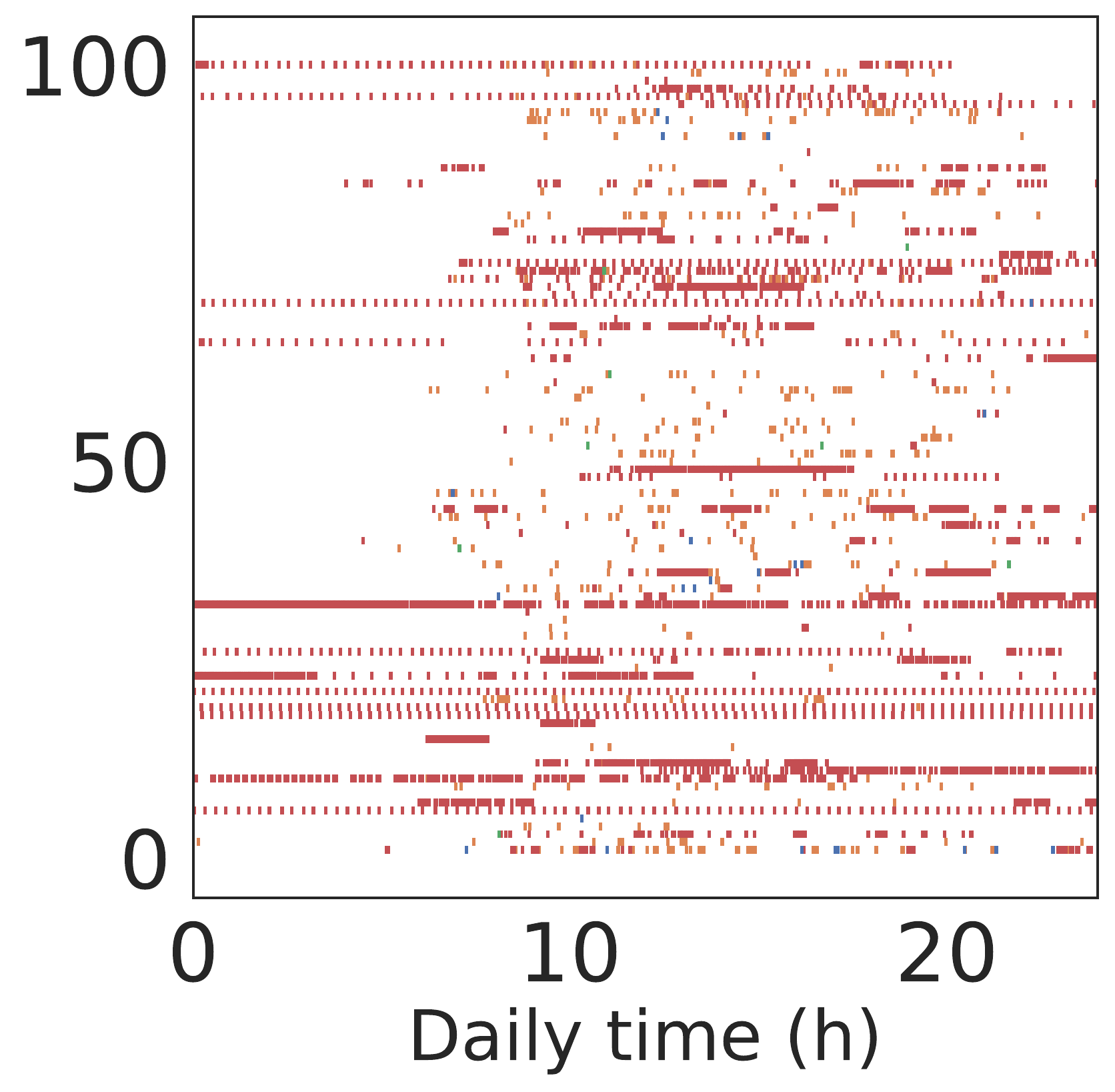}
\includegraphics[scale=0.15]{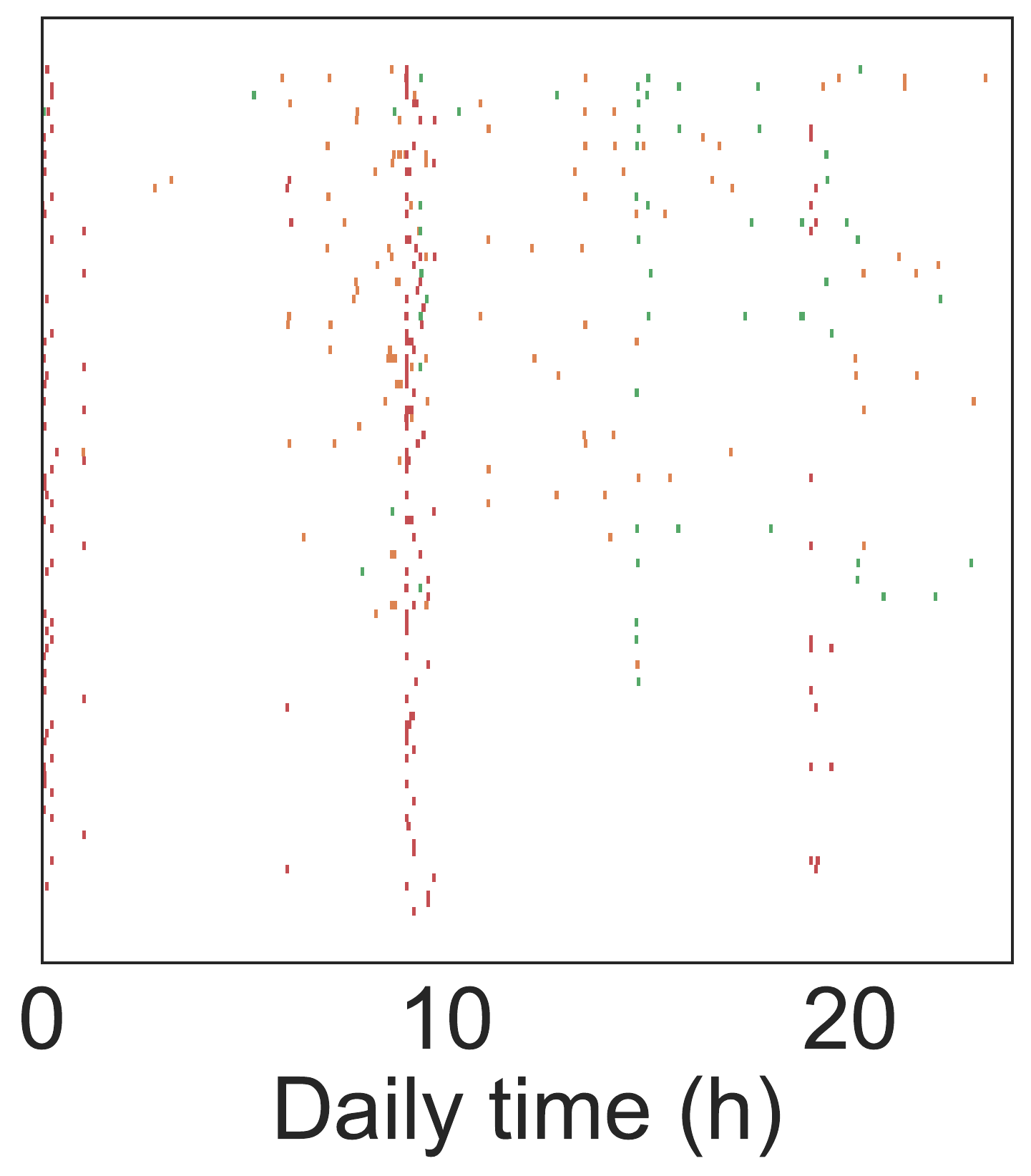}
\includegraphics[scale=0.15]{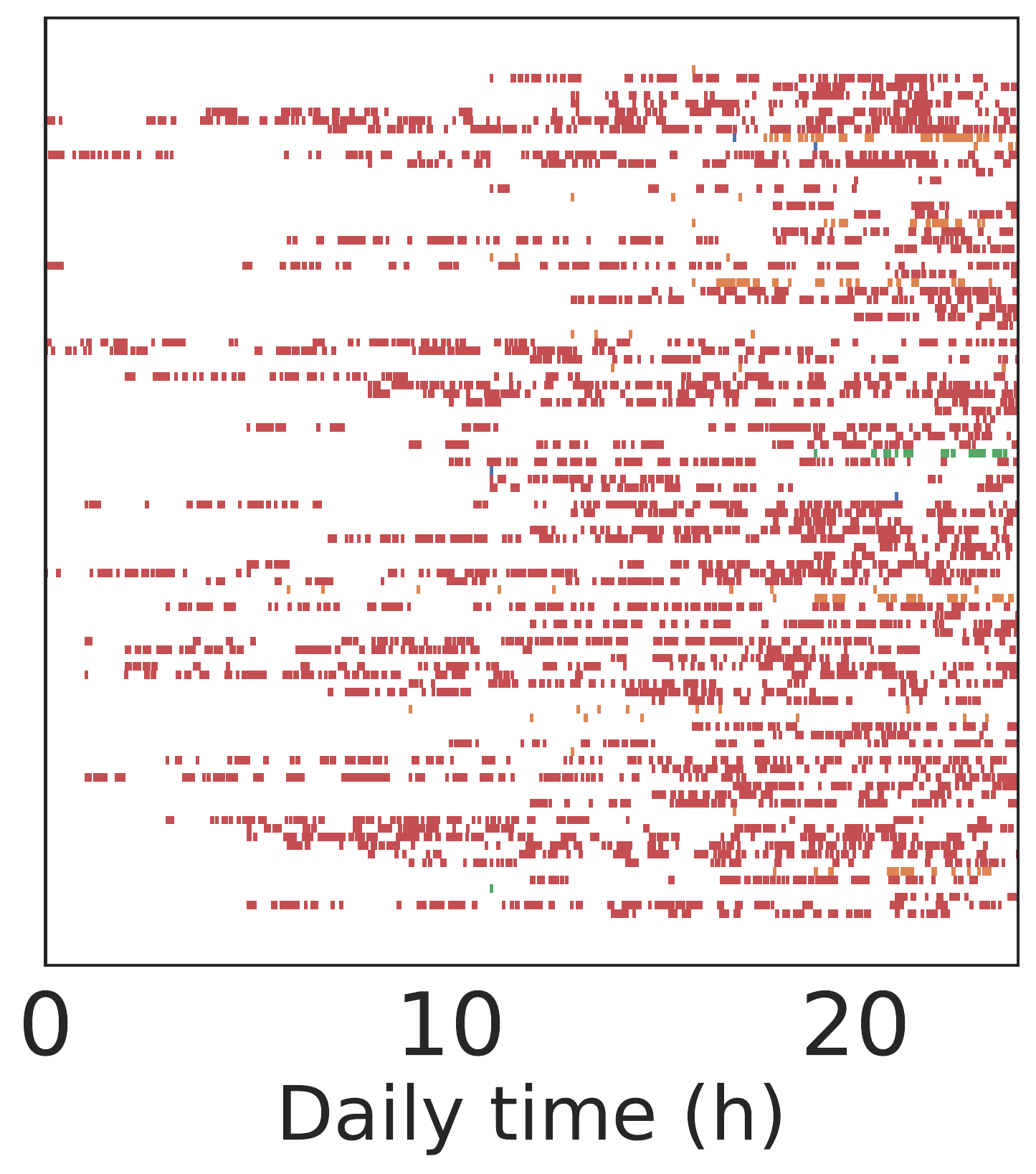}
\caption{Temporal event sequences of 100 users for: (left) a real-world, (center) a statistically- and (right) the \textit{Zen}-generated CdRs.} 
\label{fig:CdR-visual}
\end{figure}

\vspace{0.06cm}
\noindent\textbf{\textit{Fine-grained description:}}
This relates to the realistic reproduction of the individual users behaviors in terms of mobility and traffic, beyond the global behavior of the population.
In particular, daily individuals' cellular traffic presents a notable heterogeneity that challenges its reproductions. 
Fig. \ref{fig:CdR-visual}(left) shows a daily traffic (i.e., sequence of events per user) of 100 randomly selected users from a real-world CdRs. 
We can see a great diversity of users regarding events generation. 
Statistical approaches (see Fig. \ref{fig:CdR-visual}(center)), as commonly used in the state-of-the-art \cite{Murtic:2018, Milita:2021, Eduardo:2014}, are limited in reproducing such traffic dynamics as they do not allow per-user modeling but per-user profile (i.e., group of users with similar behavior). 
Improving this result, the approach we use in \textit{Zen} better captures such individuals heterogeneity (see Fig. \ref{fig:CdR-visual} right).

\vspace{0.06cm}
\noindent\textbf{\textit{Privacy compliance:}}
\textit{Zen} leverages \textit{refCdRs} with no geographical information associated with traffic events. From such CdRs, \textit{Zen} uniquely captures and reproduces individuals' traffic behavior in time, which can be then associated with any modeling of individuals' daily urban mobility. In \textit{Zen}, mobility behavior is emulated as realistically as possible. Such disjoint modeling hides real individuals' spatiotemporal daily-life habits in routine and leisure times (e.g., home/work, nightlife, etc.), bringing the privacy-preserving capability to the produced CdRs.

\begin{table}[!htb]
\begin{minipage}{.48\linewidth}
  \caption{CdRs format.}
  \label{tab:CdR}
  \centering
    \resizebox{\textwidth}{!}{%
    \begin{tabular}{@{}ll@{}}
    \textbf{}                              & \multicolumn{1}{c}{\textbf{CdR field}} \\ \midrule
    \multirow{5}{*}{\textbf{All}} & Phone number                                    \\
                                           & IMEI                                   \\
                                           & cell Id                                 \\
                                           & Timestamp                              \\
                                           & Event-type (call/SMS/data)             \\ \hline
    \multirow{4}{*}{\textbf{Call}} & Call type (MO/MT/IMO/IMT)              \\
                                           & Call duration                          \\
                                           & Phone number of the correspondent      \\ \hline
    \multirow{2}{*}{\textbf{SMS}}  & SMS type (MO/MT/IMO/IMT)               \\
                                           & Phone number of the correspondent      \\ \hline
    \textbf{Data}                  & Data session size                      \\ \bottomrule
    \end{tabular}
    }
\end{minipage}%
\hspace{1mm}
\begin{minipage}{.48\linewidth}
  \centering
    \caption{Review of generated CdRs completeness}
    \label{tab:CdR_completeness}
    \resizebox{0.78\textwidth}{!}{%
    \begin{tabular}{@{}lllll@{}}
    \toprule
    \multirow{2}{*}{\textbf{}} & \multicolumn{1}{c}{\multirow{2}{*}{\begin{tabular}[c]{@{}c@{}}Mobility\\ features\end{tabular}}} & \multicolumn{3}{l}{Traffic features} \\ \cmidrule(l){3-5} 
                 & \multicolumn{1}{c}{} & Call       & SMS        & data       \\ \midrule
    \cite{Murtic:2018}           & \xmark               & \checkmark     & \xmark     & \xmark     \\
    \cite{Milita:2021}           & \xmark               & \checkmark     &  \xmark          &    \xmark        \\
    \cite{Hughes:2019}           &  \xmark                    & \checkmark          &   \xmark         &     \xmark       \\
    \cite{Eduardo:2014}           &   \xmark                   &      \xmark      &      \xmark      & \checkmark          \\
    \cite{Gonzalo:2017}            & \checkmark                    &   \xmark         &      \xmark      &         \xmark   \\
    \cite{WHERE}           & \checkmark                    &     \xmark       &     \xmark       &        \xmark    \\
    \cite{Zilske:2014}           & \checkmark                    &    \xmark        &       \xmark     &       \xmark     \\
    \cite{Kulkarni:2017}           & \checkmark                    &    \xmark        &       \xmark     &       \xmark     \\
    \textbf{Zen} & \textbf{\cmark}           & \textbf{\cmark}  & \textbf{\cmark}  & \textbf{\cmark}  \\ \bottomrule
    \end{tabular}%
    }
\end{minipage} 
\end{table}

\section{The traffic module}
\label{sec:traffic}

We describe here the generative model used to reproduce CdRs traffic behavior.
Our generative model has enough expressive power to capture inter-CdRs feature correlations while considering individual users' behavior. In particular, we leverage an enhanced \textit{recurrent neural network} (RNN), named \textit{Long-Short-Term Memory} (LSTM), known for its ability to generate complex, realistic long-range sequences. 

Our model is trained from \textit{RefCdRs} which report a set of timestamped events generated by several users. Each CdRs' event (or a line) includes the following information: start time, user id (i.e., phone number), event-type (i.e., data, SMS, call), corresponding user id (for calls and SMS), call duration (for calls only), and data volume (for data only).

We organize \textit{RefCdRs} by user: the set of events chronologically generated by the user $u$ throughout the trace forms a sequence of events ($e^u_1, e^u_2, e^u_3, ... e^u_{N_u}$) of size $N_u$, which is the model basis. 
Hence, data reproduction is done in a sequential order, 
i.e., from time step 1 to $N_u$. The generation of an event in a sequence is a four-stage process, where each stage relies on the previous output.

\vspace{0.1cm}
\noindent\textbf{\textit{Stage 1:}} at step $t$, we predict the next event-type $e^u_{t+1}$ a user will perform, using the \textit{event-type model} (cf. \S \ref{subsec:evt}).

\vspace{0.06cm}
\noindent\textit{\textbf{Stage 2:}} given the event-type, the \textit{inter-event time (IET) model} generates the IET value used to deduce the starting time for the predicted event-type $e^u_{t+1}$ (cf. \S \ref{subsec:inter-event}).
\vspace{0.06cm}

\noindent\textit{\textbf{Stage 3:}} the \textit{correspondent model} predicts which of its correspondent a user will interact with for the next event $e^u_{t+1}$ (\S \ref{subsec:correspondent}). This model is executed only if $e^u_{t+1}$ is a call or SMS, i.e., the only events requiring correspondent interactions.

\vspace{0.06cm}
\noindent\textit{\textbf{Stage 4:}} Finally, the \textit{metric model} refers to how the events are generated: For call events, it generates its duration, while for for data events, it produces the data volume (\S \ref{subsec:metric}).
Note that the temporal information is not constant throughout the pipeline. From stages 1 to 2, we use the temporal information of the event-type at step $t$ to predict the one of the event-type at step $t+1$, then used in stage 3.

\subsection{Event-type modeling}
\label{subsec:evt}
The \textit{event-type model} predicts the next event-type a user will generate from four types of events: data, local calls (uniquely outgoing), international calls (outgoing or incoming), and local SMS (uniquely outgoing). Local incoming calls and SMS are modeled here as they are induced from outgoing calls and SMS during the generation. 
Modeling international calls separately from local calls, rather than having a unique "call" event-type and determining probalistically if it is local or international, allows distinguishing different user behaviors towards international calls. As shown in Fig. \ref{fig:CdR-visual}, some users may not make international calls while others make them frequently. 
Finally, we did not model international SMS event-type because it is rare and not present in \textit{RefCdRs}.

\vspace{0.1cm}
\noindent\textbf{\textit{The event-type model.}}
We model sequences of event-types using an LSTM.
At step $t$, the LSTM takes as input a vector of features $x_t$ and generates a vector of four scores, $y_t = (y^1_t, y^2_t, y^3_t, y^4_t)$. These scores parameterize a multinomial distribution $Pr(\widehat{e}^u_t|y_t) $ for the next event-type $\widehat{e}^u_{t+1}$, through a softmax function: 
$Pr(\widehat{e}^u_t|y_t) = \frac{exp(y^k_t)}{\sum_{k'=1}^{4} exp(y^{k'}_t)} $.\\
When training, the true previous event-types at step $t$ are encoded as input for the next step. 
Network parameters' training is done according to the standard approach of minimizing the negative-log-likelihood of the training data. We compute the gradient of this loss with respect to our network parameters through backpropagation.

\vspace{0.1cm}
\noindent\textbf{\textit{Features  $x_t$.}} 
At step $t$, we distinguish four features for predicting $e^u_{t+1}$: the event-type at step $t$ (one-hot encoded) and its temporal features, i.e., Day-of-Week (DOW, one-hot encoded), Hour-of-Day (HOD, one-hot encoded), and Second-of-Day (SOD, cyclical encoded).
A one-hot encoding represents the \textit{i}th of N features using a N-sized vector of all zeros, except for the \textit{i}th element, which is set to 1. 
A cyclical encoding maps a continuous inherently-cyclical feature into two dimensions using a sine and cosine transformation. 
The \textit{HOD} and \textit{DOW} features capture the seasonality and regularity of mobile traffic (less activity at night and during weekends~\cite{Candia:2008}). 
The fine-grained encoding of time as \textit{SOD} is used to capture the very short temporal difference between consecutive events (e.g., tens of seconds for data events).

\subsection{Inter-event time modeling}
\label{subsec:inter-event}

The \textit{IET model} returns the possible time values between a sequence's events with a confidence interval.
It works in two steps: first, we use an LSTM to parameterize a multinomial distribution over a discrete set of time bins. 
Then, we use statistical methods to sample a continuous value inside a predicted time bin.
In the following, we present our considerations for discrete IET estimation, then the detail of our LSTM network, and finally, our methodology for sampling an IET value given an IET bin.

\vspace{0.1cm}
\noindent\textbf{\textit{Discrete IET estimation.}}
IET are divided into discrete bins, $b_1,.., b_J$, representing $J$ consecutive intervals of time. 
To determine the bin boundaries, \cite{Kvamme:2021} recommends setting boundaries at evenly-spaced quantiles of time in training data. We found that, in our case, such a setting results in tiny intervals for the smallest values of IET due to the IET's heavy-tailed distribution. 
For instance, considering the 4-quantiles, there are as many elements in $[1s-20s[$ as in $[20s-72s[$. A division at the 20s could distort the model's accuracy while being acceptable for realistic CdRs.
Thus, we chose the IET bins empirically to make the model less complex and easier to train without increasing the reconstruction error in mapping back to continuous values. We, therefore, divide IET into three intervals: $[0s-30min]$  $]30min-24h]$, and $>24h$.

\vspace{0.1cm}
\noindent\textbf{\textit{The IET LSTM model.}}
The LSTM network takes at each step, $t$, as input a feature vector, $x_t$ and generates as output a vector of scores $y_t$, with one score for each possible IET bin. As with the \textit{event-type model}, these scores are used as logits in a softmax to get a multinomial distribution over the time bins. 
To train the network parameters, we minimize the negative-log-likelihood of the training data.

\vspace{0.1cm}
\noindent\textbf{\textit{Features  $x_t$.}}
At each step $t$, we consider as features, the temporal information of $e_t^u$ (\S \ref{subsec:evt}) as well as the predicted event-type $e_{t+1}^u$, one-hot encoded.

\vspace{0.1cm}
\noindent\textbf{\textit{Continuous estimation.}}
Generating CdRs traffic requires knowing the precise starting time of the next event of the sequence, which is used for further predictions. Therefore, we convert the predicted discretized IET bins to real-values. We apply to each IET bin the KS statistic test to estimate the distribution and related parameters best fitting the corresponding empirical distribution in  \textit{RefCdRs}. Table \ref{tab:iet-fitting} shows the fitted distributions to sample an IET value per bin. The model returns the median value and the confidence interval of the values obtained after $n$ sampling (by default $n =1$).

\begin{table}[]
\centering
\caption{IET distribution and parameters per bin}
\resizebox{0.45\textwidth}{!}{%
\begin{tabular}{lll}
\toprule
\textbf{IET bin}                                    & \textbf{Distribution}  & \textbf{Parameters}   \\ \hline
\multicolumn{1}{l|}{$[0s-30min]$}                 & \multicolumn{1}{l|}{Lognormal} & \multicolumn{1}{l}{$\sigma = 1.798, \mu=4.04,x_0 = 0.99$} \\ \hline
\multicolumn{1}{l|}{$]30min-24h]$}      & \multicolumn{1}{l|}{Lognormal}  & \multicolumn{1}{l}{$\sigma = 1.731, \mu=8.59, x_0 = 1749.08$}\\ \hline
\multicolumn{1}{l|}{$>24h$}                       & \multicolumn{1}{l|}{Exponential} & \multicolumn{1}{l}{$\lambda = 6.21e-06, x_0 = 86401$} \\ \bottomrule
\end{tabular}%
}
\label{tab:iet-fitting}
\end{table}


\subsection{Correspondent modeling}
\label{subsec:correspondent}
The \textit{correspondent model} applies only for event-types requiring interaction with a correspondent (i.e., SMS and local or international calls).
We first define the notion of \textit{friendship degree} ($fd$), intuitively capturing the friendship strength of a user with each of its correspondents.
Let $u$ be a user, with $\#c_u$ correspondents over the considered period, we then call $\#e^u_c$ the number of events the user $u$ had with his correspondent $c$. 
We increasingly order the correspondents of $u$ according to their corresponding number of events such that $\#e^u_1 \leq \#e^u_2 \leq .. \leq \#e^u_j \leq .. \leq \#e^u_{\#c_u}$. 
The \textit{friendship degree} of the correspondent $c$ of $u$ is the rank $j$ of $c$ in this order. 
Hence, at step $t$, the \textit{correspondent model} returns a predicted \textit{friendship degree} $\widehat{fd}_t^u$ for the correspondent  with whom the event $e^u_t$ is done.

\vspace{0.1cm}
\noindent\textbf{\textit{Correspondent LSTM model.}} 
The \textit{correspondent model} is also a LSTM network that takes as input at step $t$, a feature vector $x_t$ per user. It generates as output the predicted \textit{friendship degree} $\widehat{fd}_t^u$. The network parameters training minimizes
the Mean Absolute Error (MAE) of the training data.

\vspace{0.1cm}
\noindent\textbf{\textit{Features  $x_t$.}} 
At step $t$, the features are: the temporal information of $e_t^u$ (cf. \S \ref{subsec:evt}) except the \textit{SOD}, the  one-hot encoded event-type $e_t^u$, and the number of correspondent of $u$, $\#c_u$.
This later is constant throughout a user sequence and is essential to help the model captures that $\widehat{fd}_t^u \leq \#c_u$. 
Accordingly, it is not encoded and is left to its actual value.

\subsection{Metric modeling}
\label{subsec:metric}
This section presents the models used to generate the metrics (i.e., a model per metric) associated with events generation, namely the call duration and the data volume. 

\begin{figure}
\begin{subfigure}{0.23\textwidth}
\includegraphics[scale=0.40]{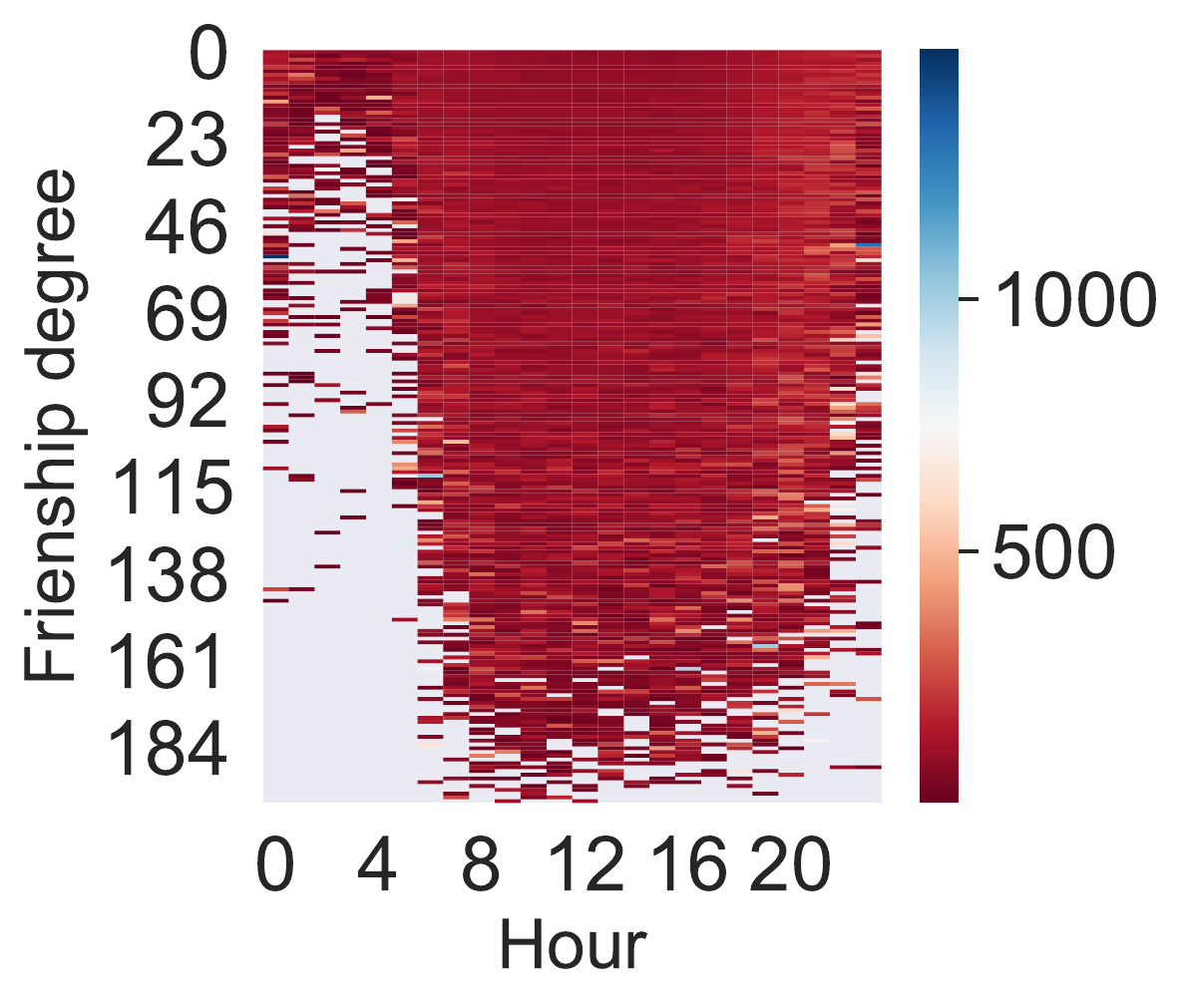}
\caption{} 
\label{fig:avg_call_duration}
\end{subfigure}
 \hfill
\begin{subfigure}{0.23\textwidth}
\includegraphics[scale=0.40]{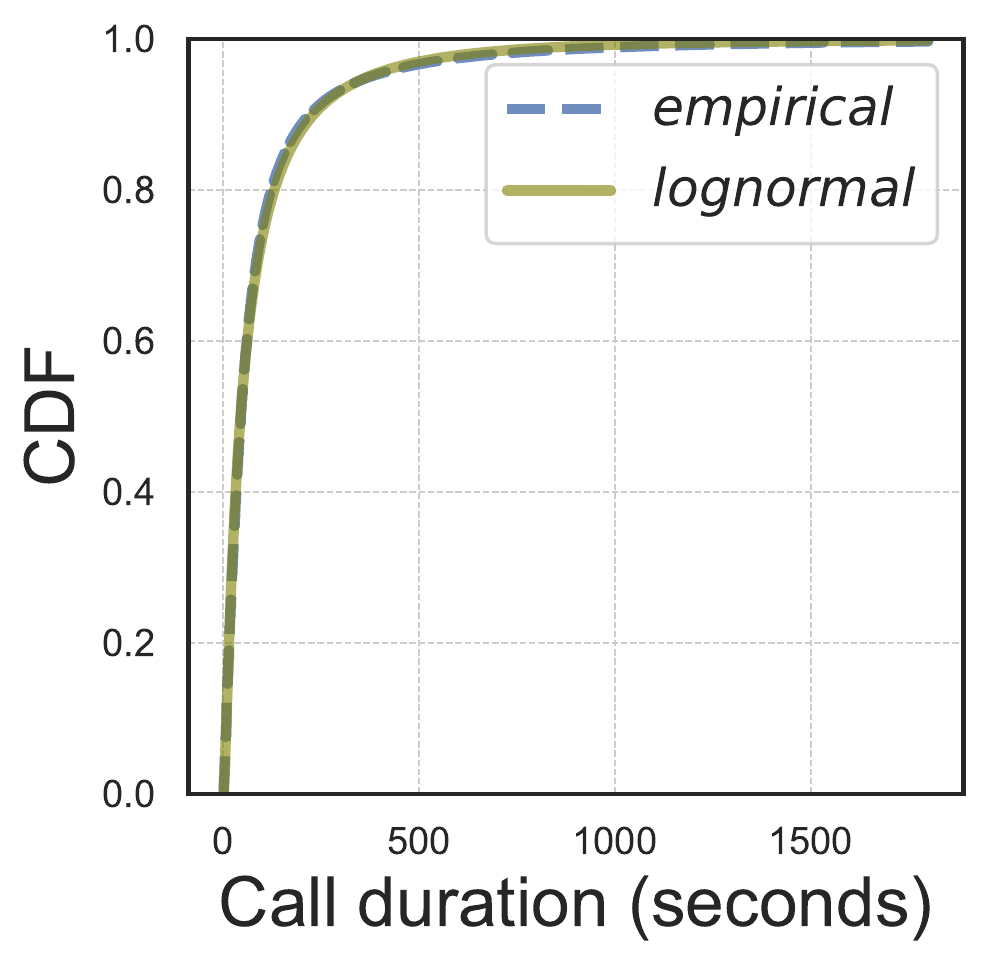}
\caption{} 
\label{fig:call_duration_distrib}
\end{subfigure}
\caption{(a) Avg call duration (s) per hour and friendship degree (b) Call duration CDF for \textit{RefCdRs}. 
}
\label{fig:call_duration_model} 
\end{figure}

\vspace{0.1cm}
\noindent\textbf{\textit{Call duration}}
We use a statistical method to model the call duration. 
In fact, contrary to the previously modeled parameters, there is no explicit features dependency or variability (and therefore, no complexity) regarding call durations, which implies that a used RNN could hardly train. This is confirmed in Fig. \ref{fig:avg_call_duration}, which shows the variation of the average call duration per hour and per friendship degree over the entire dataset. We can see that overall, call duration does not vary much, and thus, there is no particular correlation between these parameters. 
Moreover, the per-user behavior regarding call duration (easily assessed through the average call duration per user) closely depends on the number of calls each user makes over the CdRs duration, which is opportunely already captured by the \textit{IET model}.
Accordingly, the \textit{call duration model} corresponds to the estimation of the parameters of the continuous distribution that best fits the empirical distribution of call duration, as shown in Fig. \ref{fig:call_duration_distrib}. From a statistical test, we found this distribution to be Lognormal of parameters $\sigma = 1.29, \mu= 3.78, x_0=-0.47$. 

\vspace{0.1cm}
\noindent\textbf{\textit{Data volume.}}
The \textit{data volume model} returns a data volume value for each data event. According to 3GPP standards, each data-typed CdRs line corresponds to the generation of a data session by a user. 
Unfortunately, as \textit{RefCdRs} lack this information, we rely on the study done in \cite{Eduardo:2014} to design the \textit{data volume model}. 
To the best of our knowledge, \cite{Eduardo:2014} is the only work 
that conducts a thorough characterization of data volume usage per session and per user over time extracted from real-world CdRs, as well as designs a generator of realistic CdRs that conforms to these characterizations. \\
\cite{Eduardo:2014} profiled users' data usage over time according to their generated amount of data (\textit{volume profile}, i.e., Light, Medium, or Heavy) and to how often they generate data sessions (\textit{frequency profile}, i.e., Occasional or Frequent). Besides, it extracted from real-world CdRs the distributions of data session volume according to a user's profile and the day period (peak or off-peak hours) and the percentage of users per profile.
We use such percentages to first assign a \textit{volume profile} to each user in \textit{Zen}.
As the \textit{frequency profile} could be inconsistent with the frequency of data event-type as predicted by the \textit{event-type model}, we attribute to each user, in \textit{Zen} the \textit{Occasional frequency profile}. In fact, the distribution of the number of data sessions per day and user from \textit{RefCdRs} shows the majority of the population to be of this latter profile. 
Finally, we sample from the distributions found in \cite{Eduardo:2014} to get a data session volume.

\section{The mobility module}
\label{sec:mobility}

The \textit{Zen's mobility module} produces realistic CdRs mobility traces in three steps each covered by a sub-module. 
The \textit{mobility-generator} (\S \ref{subsec:mobility-generator}) emulates a population urban mobility in a real-world city map, with users displacements generated according to public sources \cite{OpenStreetMap} and describing city planning and transportation information~\cite{HelsinkiReport}. 
Next, from the input city map, the \textit{topology-builder} (\S \ref{subsec:topology-builder}) builds a realistic cellular 
topology using cell towers' positioning of mobile operators deployed in the considered real-world city, gotten from OpenCellID \cite{OpenCellID}.
This topology is then used in the \textit{position-to-cellId module} (\S \ref{subsec:combining-module}) to map the mobility traces produced by the \textit{mobility-generator} to the cell granularity.

\subsection{Mobility generator}
\label{subsec:mobility-generator}

\textit{Zen mobility-generator} inherits the highly configurable capability of the \textit{Opportunistic Network Environment} (ONE)~\cite{ONE} simulator.
Besides, it enhances the \textit{Working Day Mobility} model (WDM)~\cite{WDM} of ONE into a model named \textit{En-WDM}, and generates CdRs of format \verb|<Timestamp|, \verb|userId|, \verb|lat,lon>|. 

Our motivation to use WDM as a basic building block of \textit{Zen} is twofold. First, contrary to related models~\cite{map-slaw, StreetModel}, WDM originality comes from the combination of various mobility aspects present in people daily life (e.g., home and workplaces, day periods). Second, WDM closely reproduces wireless interactions (i.e., inter-contact and contact time) distributions found in two real-world measurement experiments (i.e., iMote and Dartmouth), asserting modeling generality.\\
Nevertheless, WDM is limited in capturing some fine-grained real mobility habits or fine-tuning. \textit{En-WDM} tackles such limitations and strengthens the model with additional literature's intuitions on laws dictating human mobility behavior, such as preferential attachment, regular daily behavior, transportation-dependent shortest-path preferences,
and most importantly, uncertainty (i.e., novelty-seeking behaviors) and heterogeneity.
Next, we detail \textit{En-WDM}.

\vspace{0.1cm}
\noindent\textbf{\textit{WDM's inherited functionalities.}} 
\textit{En-WDM} models week working days' movements into three activities and their transitions, i.e., ''home'', ''working'', and ''night activity''. The night activity corresponds to leisure-related times spent in preferred spots of friends groups.

\vspace{0.1cm}
\noindent\textbf{\textit{Exploration profiling.}}
Users in \textit{En-WDM} emulation decide in a probabilistic-way whether to go home or to a night activity. To setup such probabilities, we rely on the exploration phenomenon profiling conducted in \cite{Amichi:2020} and define three mobility profiles: \textit{scouters} are more inclined to explore and discover new places to visit, \textit{routiners} rarely explore and prefer to stay among their familiar and few known places, and \textit{regulars} constantly alternate between exploration and routine. We then accordingly classify users given by the \textit{ChineseDB} dataset (cf. Sec \ref{subsec:datasets}) in these three profiles.
Results describe a population with 20.27\% of \textit{scouters}, 54.75\% of \textit{regulars}, and 24.98\% of \textit{routiners}. After this classification, we assign to users in each profile, the probabilities of ``nightlife activity'': 0.8 for \textit{scouters}, 0.5 for \textit{regulars}, and 0.2 for \textit{routiners}. 
    
\vspace{0.1cm}
\noindent\textbf{\textit{Neighborhood and popularity.}} 
Rather than considering home/office locations' (lat, lon) coordinates, \textit{En-WDM} associates each location coordinate to the center of a neighborhood of rectangular shape and configurable size.
This allows to distinguish areas with high housing density (e.g., residential areas, university campuses), areas with high business density (business districts), and popular leisure locations.
A user is first assigned a home/office neighborhood and then, chooses her exact home/office location randomly inside the neighborhood. 
Moreover, we added the \textit{neighborhood popularity}, which represents the probability for a user to choose a given neighborhood as a home/office neighborhood or, in the case of night activity, the probability of choosing a spot for her evening activity.


\vspace{0.1cm}
\noindent\textbf{\textit{Distance-based profiling.}}
\textit{En-WDM} enables the definition of cities' 
districts (hereafter, areas) to replicate the real world. Accordingly, we associate each user to one of the three profiles representing area displacements: \textit{profile 1} inside a single area, \textit{profile 2} among two areas, and \textit{profile 3} in the whole map.
To get the population percentage to be considered in each profile, we profile \textit{Geolife}'s users resulting in: \textit{Profile 1} including 72\% of users whose maximum distance is less than 1/3 of the maximum observed distance $D_{max}$ ($\approx \num{2.49e3}km$). \textit{Profile 2} with 19\% of users with a maximum distance between 1/3 and 2/3 of $D_{max}$, and \textit{profile 3} including 9\% of users with a maximum distance greater than 2/3 of $D_{max}$.

\vspace{0.1cm}
\noindent\textbf{\textit{Simple parameterization.}} 
We report here all the key configuration parameters needed for \textit{En-WDM} emulation. Table \ref{tab:wdm_parameterization} summarizes them. We use italic style for those we used the default value and regular one for those we modified.
We set the value of \textit{ProbOwnCar} to 0.19 based on transportation statistics in the city of Helsinki~\cite{HelsinkiReport}.
Parameters in bold (\textit{homeRange} and \textit{officeRange}) are those we added for clusters implementation. 
In particular, the ratio between the \textit{worldSize}, \textit{officeSize}, and cluster sizes may vary depending on the emulated city. These parameters values in Table \ref{tab:wdm_parameterization} are adapted for a emulation in the city of Helsinki. 

\begin{table}[]
\centering
\caption{Key parameters for \textit{En-WDM} emulation}
\label{tab:wdm_parameterization}
\resizebox{0.47\textwidth}{!}{%
\begin{tabular}{ll}
\hline
\textbf{\textit{En-WDM} Parameter Description}                                                & \textbf{Value vs Default} \\ \hline
\multicolumn{1}{l|}{\textit{Size of the office squared-shaped side}}                 & \textit{100}              \\ \hline
\multicolumn{1}{l|}{\textit{Minimum size of a friends group for evening activities}} & \textit{1}                \\ \hline
\multicolumn{1}{l|}{Maximum size of a friends group for evening activities}          & 5 vs 3                    \\ \hline
\multicolumn{1}{l|}{Minimum value for evening activities duration}                   & 1h vs 10s                 \\ \hline
\multicolumn{1}{l|}{Maximum value for evening activities duration}                   & 4h vs 2h                  \\ \hline
\multicolumn{1}{l|}{Probability for a user to own a car}                             & 0.19 vs 0.5               \\ \hline
\multicolumn{1}{l|}{\textit{(width, heigth) of the emulation area}}                 & \textit{(10000, 8000)}    \\ \hline
\multicolumn{1}{l|}{\textbf{(width, heigth) of a home cluster}}                      & \textbf{(250, 150)}       \\ \hline
\multicolumn{1}{l|}{\textbf{(width, heigth) of a office cluster}}                    & \textbf{(500, 300)}       \\ \hline
\end{tabular}%
}
\end{table}

\subsection{Topology builder}
\label{subsec:topology-builder}

The \textit{topology-builder} uses the geographical positions of base stations (BS) in the emulated area, as given by OpenCellId~\cite{OpenCellID}, and performs a Voronoi tesselation. The tessellation produces a cellular network topology with heterogeneous cell sizes close to reality, containing each input BS. Each Voronoi cell defines the communication boundaries of an input BS.
For generality and simplicity reasons, we include all operators' base stations given by OpenCellId in a bigger architecture to derive the Voronoi topology. This unique topology is assigned to all operators considered in \textit{Zen}'s process. In practice, sharing BSs between different operators is commonly done for cost savings.


\subsection{Position-to-cellId module}
\label{subsec:combining-module}

The \textit{position-to-cellId module} assembles the modeled users' mobility and the designed Voronoi cellular topology. For this, each user's geographical position given by the \textit{mobility-generator} traces is mapped to the corresponding OpenCellId's BS identifier, i.e., cellID, in the Voronoi topology. It outputs mobility CdRs in the format \verb|<Timestamp|, \verb|userID|, \verb|cellID>| describing users' spatiotemporal daily mobility in a real city map and adapted to a real network topology. Despite the realism given by such leveraged real-world information, the generation of users' mobility has a realistic and not a real nature since no ground-truth information on users' real-life routine is available. This brings privacy benefits to \textit{Zen}  CdRs.

\section{The \textit{social-ties} module}
\label{sec:social_ties}

\textit{Zen} CdRs generation lays on the \textit{social-ties module} providing the network social structure. 
This structure induces phone numbers from users of the mobility CdRs and builds the network social graph by creating per user's phonebook, i.e., the users she can interacts through calls or SMS.



\vspace{0.1cm}
\noindent\textbf{\textit{Mobility users to phone numbers.}}
From the number of network's operators and the users distribution per operator (taken as parameter or induced from OpenCellId\cite{OpenCellID}), the \textit{social-ties module} assigns an operator per user and generates a 
phone number in the format \verb|<MCC><MNC>| \verb|<5 random| \verb| digits>|, where MCC and MNC describe the mobile code for country and the operator network code within the country, respectively.

\vspace{0.1cm}
\noindent\textbf{\textit{Network social graph.}}
Reproducing the social graph of users' interactions implies answering the following three questions.
The term "correspondent" refers to a phone number in a user's phonebook. 

\vspace{0.1cm}
\noindent\textbf{(Q1) how many correspondents does each user have? }
To answer this question, the \textit{social-ties} module relies on the distribution of correspondents per user from \textit{RefCdRs}.
Let $u \in U$ be a user with $\#c_u$ correspondents; we consider the non-parametric distribution $P_{\#c} = P(\#c_u = \#c) \;\; \forall \; \#c \in [1, MAX]$.
Thus, for each generated user $u'$, its number of correspondents $\#c_{u'}$ is obtained with the multinomial distribution of parameters $P_{\#c}$.\\
We then define four disjoint categories of correspondents: international correspondents ($c_{inter}$), outgoing local correspondents ($c_{out}$), incoming local correspondents ($c_{in}$), and both outgoing and incoming local correspondents ($c_{both}$).
Thus, $\forall \; u \in U, \; \#c_u = \#c_{inter,u} + \#c_{out,u} + \#c_{in,u}+ \#c_{both,u} = (x_{inter,u} + x_{out,u} + x_{in,u}+ x_{both,u})\times\#c_u$.
We export the average values $\overline{x_{cat,u}} \;\; \forall \; cat \in \{inter,\; out,\; in,\; both\}$. 
Then, we use the multinomial distribution of $P = \overline{x_{cat,u}}$ to induce the number of correspondents, in each category, of each user.

\vspace{0.1cm}
\noindent\textbf{(Q2) how do we choose these correspondents?}
We create user phonebooks by implementing a variant of the configuration model algorithm~\cite{wiki:ConfigurationModel}, which allows building a graph from given users degrees. We apply this algorithm by correspondents' category so that each user is an $c_{in}$ correspondent of its $c_{out}$ correspondents and a $c_{both}$ of its $c_{both}$ correspondents.
Moreover, we add a heuristic to choose users' correspondents based on their relationship type, (i.e., neighbors, colleagues, or friends) extracted from the generated mobility dataset (cf. \S \ref{sec:mobility}) as follows.
users located inside the same home/work cluster between \textit{1am to 4 am} and \textit{10 am to 2 pm}, over the whole dataset duration, are considered neighbors and colleagues, respectively. As well, users in the same group for night activities, when they occur, are considered friends. 
Hence, a user's correspondents are selected according to defined probabilities (taken as parameters) from its list of neighbors, colleagues, friends, and other users until we reach the fixed number of the user's correspondents.

At last, the \textit{social-ties} module outputs each user's list of correspondents organized in the categories $c_{out}$, $c_{both}$, and $c_{inter}$, while $c_{in}$ category is induced from the $c_{out}$ one.

\vspace{0.1cm}
\noindent\textbf{(Q3) how does a user interact with all of its correspondents?}
While (Q1) and (Q2) are tackled by the \textit{social-ties module}, question (Q3) is addressed through the \textit{correspondent model} of the \textit{traffic module} detailed in section \ref{subsec:correspondent}.

\section{The \textit{Cdr-combiner} module}
\label{sec:zen-merger}
\textit{Zen}'s \textit{CdR-combiner} module integrates outputs of previous modules to produce realistic CdRs, as follows.

Using \textit{event-type} and \textit{IET models} from the \textit{traffic module}, the \textit{CdR-combiner} generates timestamped sequences of events over the total duration.
Then, each sequence is associated with a correspondent determined by the 
\textit{social-ties} module, based on each user's number of correspondents per category, indicating which event-types the user can generate. 
At this point, using the \textit{correspondent model}, the \textit{CdR-combiner} predicts a correspondent friendship degree per user event that is later associated with the corresponding phone number from users' phonebooks.

Next, we add complementary metrics to users' events. 
For all calls events, the call duration metric relates only to available correspondents of users. We do not consider unavailable users' correspondents (i.e., already in an ongoing communication) at the caller-callee association. Hence, for available correspondents, a call duration value is sampled from the \textit{call duration} model distribution. This value is upper-bounded by the time to the closest scheduled call. 
As well, for data events, the data volume metric is assigned according to the \textit{data volume} model. 

Following, the \textit{CdR-combiner} integrates CdRs spatial information, i.e., corresponding users' cell Ids at each event timestamp (resulting from the \textit{mobility module}). At last, based on users' phone numbers, the \textit{CdR-combiner} infers CdRs traces produced by each operator. \textit{Zen}, therefore, generates a complete and realistic CdRs trace per generated mobile operator in the format specified in Table \ref{tab:CdR}.

\section{Evaluations}
\label{sec:evaluation}
This section confirms \textit{Zen}'s validity by evaluating traffic and mobility modeling separately, then their merging into CdRs.

\subsection{Traffic module}

Hereafter, we evaluate the accuracy and the performance of predictions resulting from the \textit{traffic module}'s stages.
As there is no similar contribution in the literature, we compare \textit{Zen}'s models to designed baseline predictors.
Table \ref{tab:lstm_evaluation} summarizes all comparison metrics and provides their distributions on the right of each evaluation result. 

\subsubsection{\textbf{Experimental datasets}}
\label{subsubsec:datasets}


We train and evaluate our models on \textit{RefCdRs} after some data handling. First, we only consider events of \textit{users subscribed} to the operator network collecting \textit{RefCdRs}. 
Then, we filter out users having less than 3 generated events in the whole period of 4 weeks and those with more than one event at the same timestamp. Those manipulations result in the selection of nearly 6000 users totalizing 1,782,829 events or CdRs entries, i.e., 77.8\% of the \textit{RefCdRs}' initial size.
We then use as \textit{training set} the first two weeks of the dataset, 
 the 3rd week as \textit{validation set}, and the 4th week as the \textit{test set}.
Because our traffic predictions are user-based, the non-filtered remaining users compose all the three previous sets and only their event sequences varies according to the week considered in each set.

\subsubsection{\textbf{Models training and Hyper-parameters}}
We used a 2-layer LSTM with 50 hidden units per layer for the \textit{event-type model} and 100 hidden units per layer for the two other models.
To avoid over-fitting the training dataset, we used a dropout regularization with $p=0.2$.
The LSTM losses are iteratively minimized using mini-batch gradient descent with the Adam optimizer. Each mini-batch contains 64 sequences of events (i.e., users). 
We chose event sequences' lengths of 302 for training, 157 for validation, 159 for test,  sampled from the distribution of the number of events generated by users in each experimental set.
Therefore, we pad all sequences to the sequence length in each experimental set to homogenize datasets and ease the training. We use a masking layer to tag added values in each sequence to ignore them in the loss calculation. Besides, we fixed a gradient clip value of $0.01$ to avoid "exploding gradients" prone to affect RNN.

\subsubsection{\textbf{Event-type model}}
We compare our \textit{event-type model}'s predictions (cf. \S \ref{subsec:evt}) to the ones of the following baselines:
\textit{Uniform}~-- each event-type is equally likely to occur at each time step; \textit{Multinomial}~-- each event-type probability is given by its empirical count 
in training data; \textit{RepeatEvt}~-- the next event-type is always predicted to be the same as the previous one.
We use the following evaluation metrics:
(\textit{NLL}) Negative-log-likelihood of next-step probabilities, and (\textit{Accuracy}) next-step 1-best correct classification rate (for this metric, the traditional Multinomial approach always output the most frequent event-type).
Results are presented in Table \ref{tab:lstm_evaluation}. Selecting event-type according to the \textit{Multinomial} is significantly more predictive than the \textit{Uniform}, but worse than \textit{RepeatEvt}. Our \textit{Zen}'s \textit{event-type model} works the best. For both NLL and Accuracy, \textit{Zen} is significantly better than \textit{RepeatEvt}, i.e., the most probable event-type is not always the previous one.

\begin{table*}
  \caption{\textit{Traffic LSTM models evaluation results.}} 
  \label{tab:lstm_evaluation}
  \includegraphics[width=0.95\linewidth]{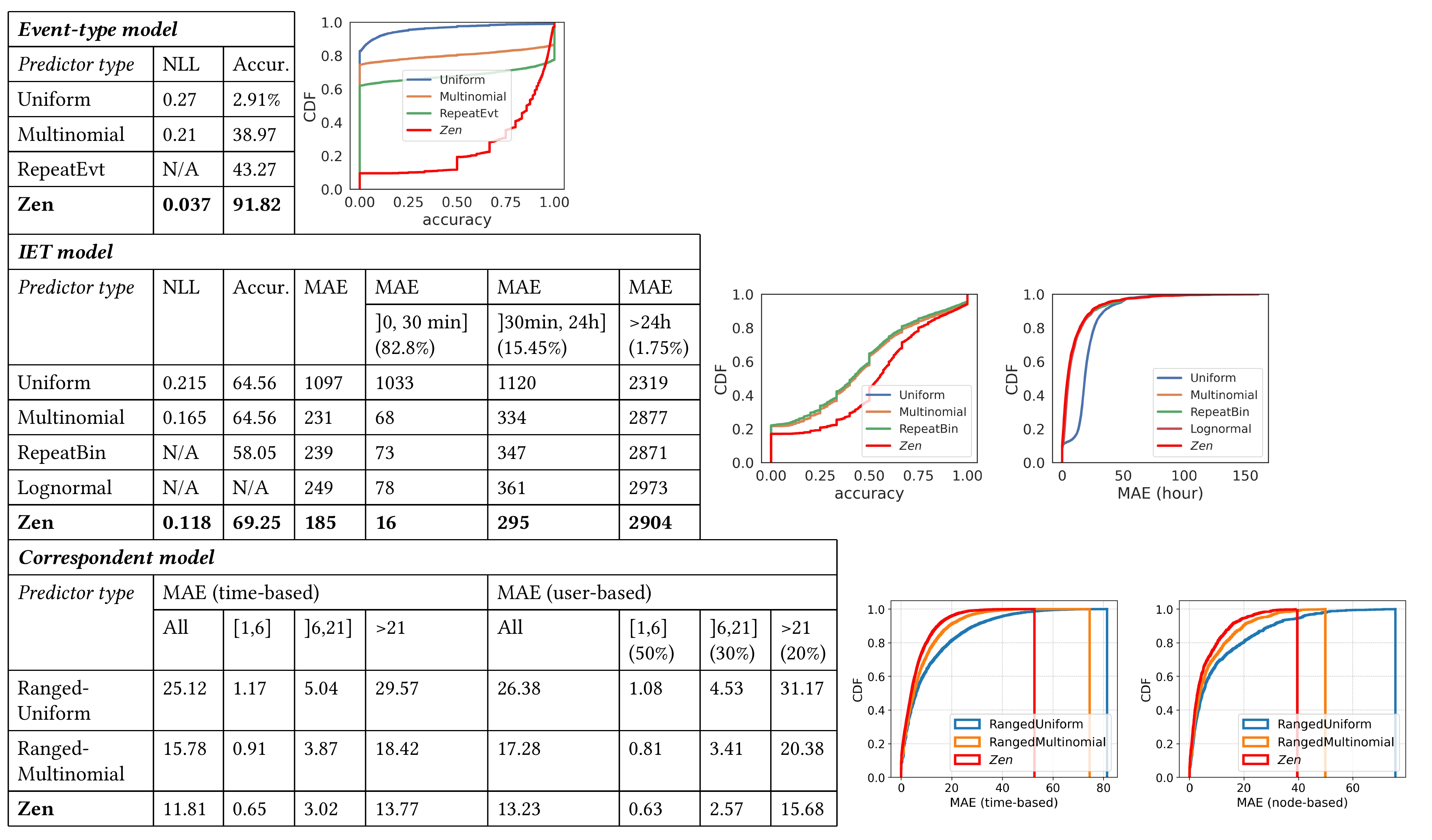}
\end{table*}

\subsubsection{\textbf{IET model}}

As before, we compare the acuteness of our model in predicting the next IET Bin (cf. \S \ref{subsec:inter-event}) with the corresponding above-defined baselines.
Table \ref{tab:lstm_evaluation} shows that for both metrics, NLL and Accuracy, the performance of \textit{Zen}'s \textit{IET model} is much higher than \textit{RepeatBin} (that simply repeats the previous IET Bin), followed by the Uniform and the \textit{Multinomial} baselines.
Disregarding the prediction approach, we compute the discretized probabilities of IET Bins and map them to IET values in a continuous domain: named \textit{Bin sampling} mapping. 
To evaluate how efficient \textit{Zen}'s and baselines' \textit{Bin sampling} are, we compare them to the \textit{Overall sampling} mapping, both described next.
 
\vspace{-0.05cm}
\begin{enumerate}[label={},leftmargin=*]
    \item[-] \textit{Bin sampling}: At each Bin,
    the IET value is obtained after averaging $n=500$ samplings of the corresponding continuous IET distribution (see \S \ref{subsec:inter-event}). We apply this approach to all the previously Bin-based models, i.e., \textit{Zen}'s \textit{IET model},
    Uniform, Multinomial, and RepeatBin predictors.
    \item[-] \textit{Overall sampling}: We perform a fitting of the empirical IET distribution (i.e., with no bins) and obtain a Lognormal distribution with 
    $\sigma=2.67, \mu=4.97, x_0=1$. Then, we straightly predict continuous values by sampling the resulted fitted IET distribution. We name this prediction \textit{Lognormal}.
\end{enumerate}
 The Mean Absolute Error (MAE) of the IETs in minutes is used as the comparison metric. It estimates the average distance between actual and predicted IET.
From Table \ref{tab:lstm_evaluation}, we can notice that the \textit{Bin-sampling} of \textit{Multinomial} and \textit{RepeatBin} have comparable MAE performances, followed by the \textit{Overall-sampling} \textit{Lognormal} predictor. This behavior is also verified per Bin (three last columns).
Overall, \textit{Zen} works the best. In the first bin $]0, 30min]$, which is the most sensitive, we note that except for the \textit{Zen}, all models on average predict an IET value outside the initial interval.

\subsubsection{\textbf{Correspondent model}}

At last, we evaluate the \textit{correspondent model} (cf. \S \ref{subsec:correspondent}) by comparing its predictions to the following baselines:
\begin{itemize}[label={},leftmargin=*]
    \item[-] \textit{RangedUniform}: Per user $u$, 
     correspondents $c_i, \; \forall i = 1,2,...,\#c_u$ are equally likely to be predicted at each sequence step. 
    \item[-] \textit{RangedMultinomial}: 
    Per user \textit{u}, 
    each correspondent $c_i$ is 
    chosen with a probability ($p^u_i, 1\leq i \leq\#c_u$) extracted from the procedure as follows:\\
    Let $U$ be the set of users and $u$ a user in $U$. We recall that $\#e^u_{c_i}$ refers to the number of events $u$ made with his correspondent $c_i$. From this definition, we derive $P^u_{c_i}$ the proportion of events made by $u$ with its correspondent $c_i$ :
    $P^u_{c_i} = \#e^u_{c_i}/\sum_{i} \#e^u_{c_i}$.\\
    For all $i = 1,2,..., MAX(\#c_u)$ we extract the mean values $\overline{P_{c_i}} = \overline{P^u_{c_i}} \; \forall u \in U$.
    Hence, for a user $u$, the probabilities ($p^u_i, i=1,2,..\#c_u$) is obtained by normalizing the first $\#c_u$ mean values ($\overline{P_{c_i}}, i=1,2,..\#c_u$) such that $\sum_{i} p^u_i = 1$.
 
\end{itemize}
The evaluation metric is the MAE of the predictions $\widehat{fd}_t$ in the test dataset. 
We found that as we train the \textit{correspondent model} with chronologically-separated experimental windows (defined in \S \ref{subsubsec:datasets}), the MAE loss value continually increases in the validation dataset. This is due to the fact that in the training period (i.e., first two weeks), users only interact with some of their correspondents, making it difficult for the model to generalize.
To fix this issue, we instead split training, validation, and test datasets by selecting users traffic over the whole dataset period (4 weeks). The training dataset includes 60\% of the users, while the validation and test datasets each represent 20\%. 
Results in Table \ref{tab:lstm_evaluation} show
the \textit{RangedMultinomial} predictor has significantly better results compared to the \textit{RangedUniform} predictor. 

Overall, \textit{Zen} is the modeling that best performs, showing its ability to capture users interaction with their correspondents. 
In particular, the detailed distribution plots show \textit{Zen} presents for 80\% of users (i) more than 95\% and 75\% of accuracy for respectively, the event-type and IET models, and (ii) less than 6.68\% and 12.5\% of MAE maximum values for respectively, the IET and correspondent models.

\subsection{Mobility module}

We validate our \textit{En-WDM} mobility model by comparing it to its original version, the WDM~\cite{WDM}. We rely on WDM results closely following real-world measurement datasets distributions (i.e., iMote or Dartmouth). Since \textit{En-WDM} adds new functionalities in modeling mobility to WDM, we are not looking for identical results from both models but for similarities in terms of distributions and curve behaviors.

\begin{figure*}
\begin{subfigure}{0.19\textwidth}
\includegraphics[scale=0.28]{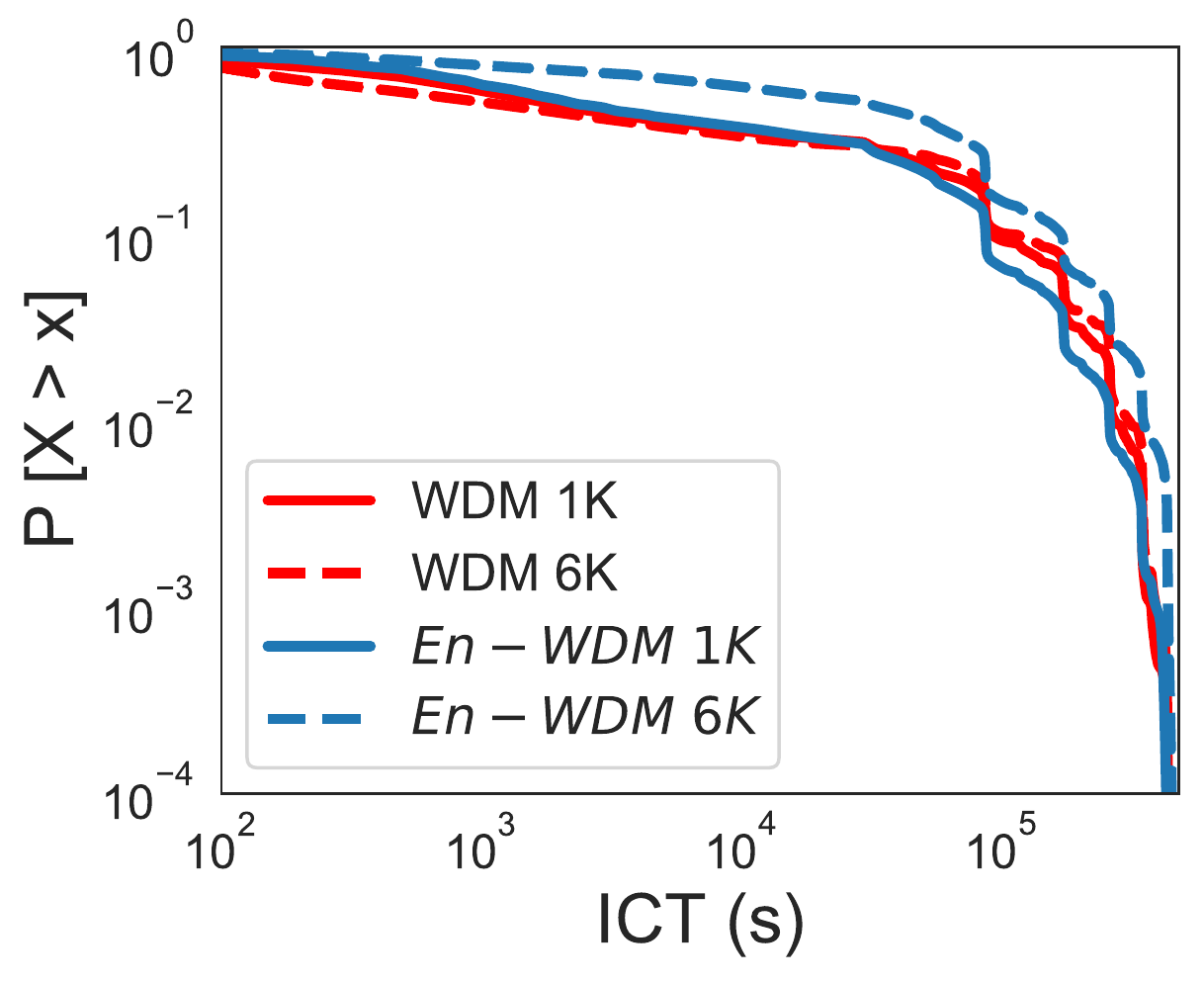}
\caption{} 
\label{fig:inter-contact-ccdf}
\end{subfigure}
\begin{subfigure}{0.19\textwidth}
\includegraphics[scale=0.28]{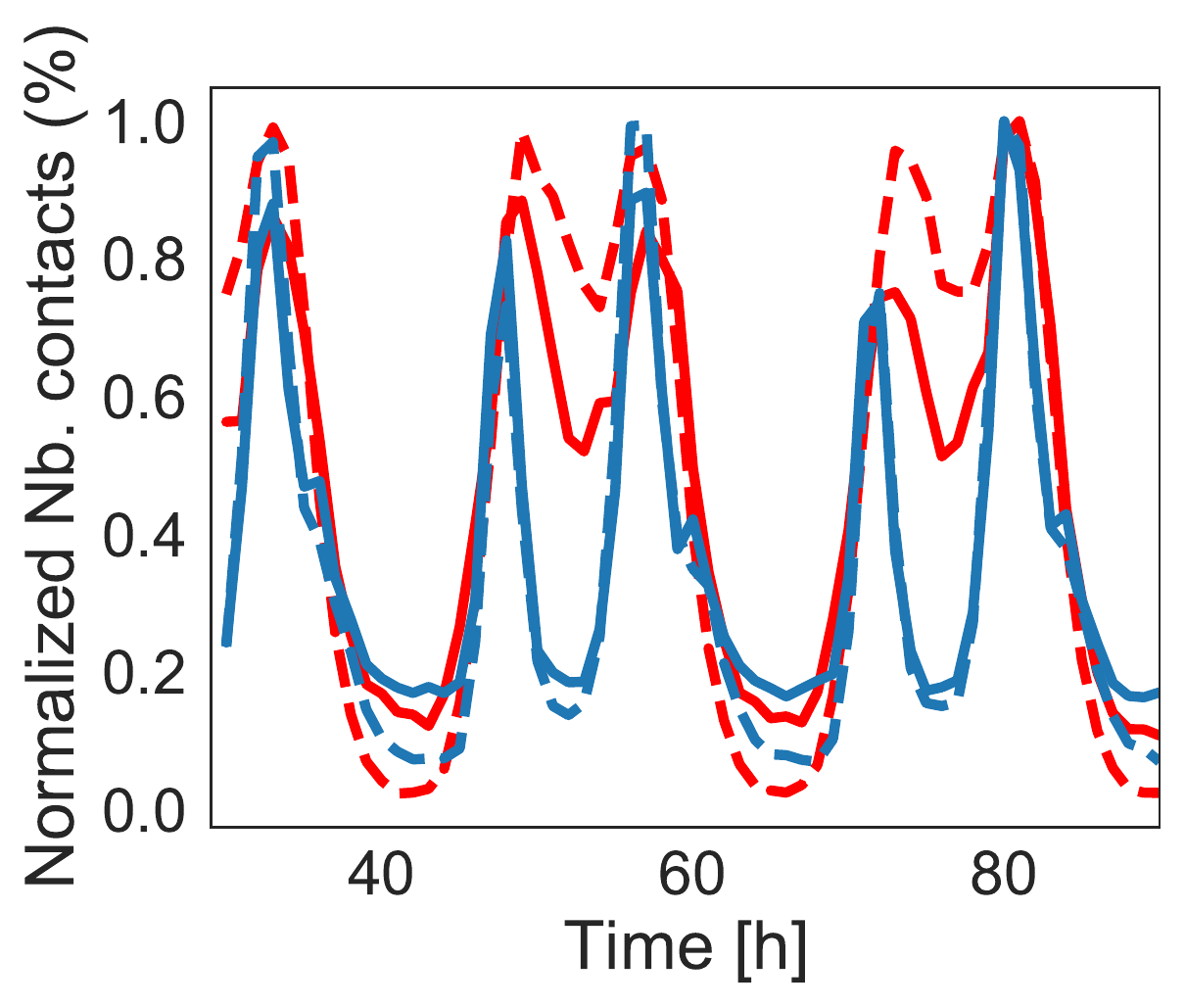}
\caption{} 
\label{fig:contact-per-hour-ccdf}
\end{subfigure}
\begin{subfigure}{0.19\textwidth}
\includegraphics[scale=0.28]{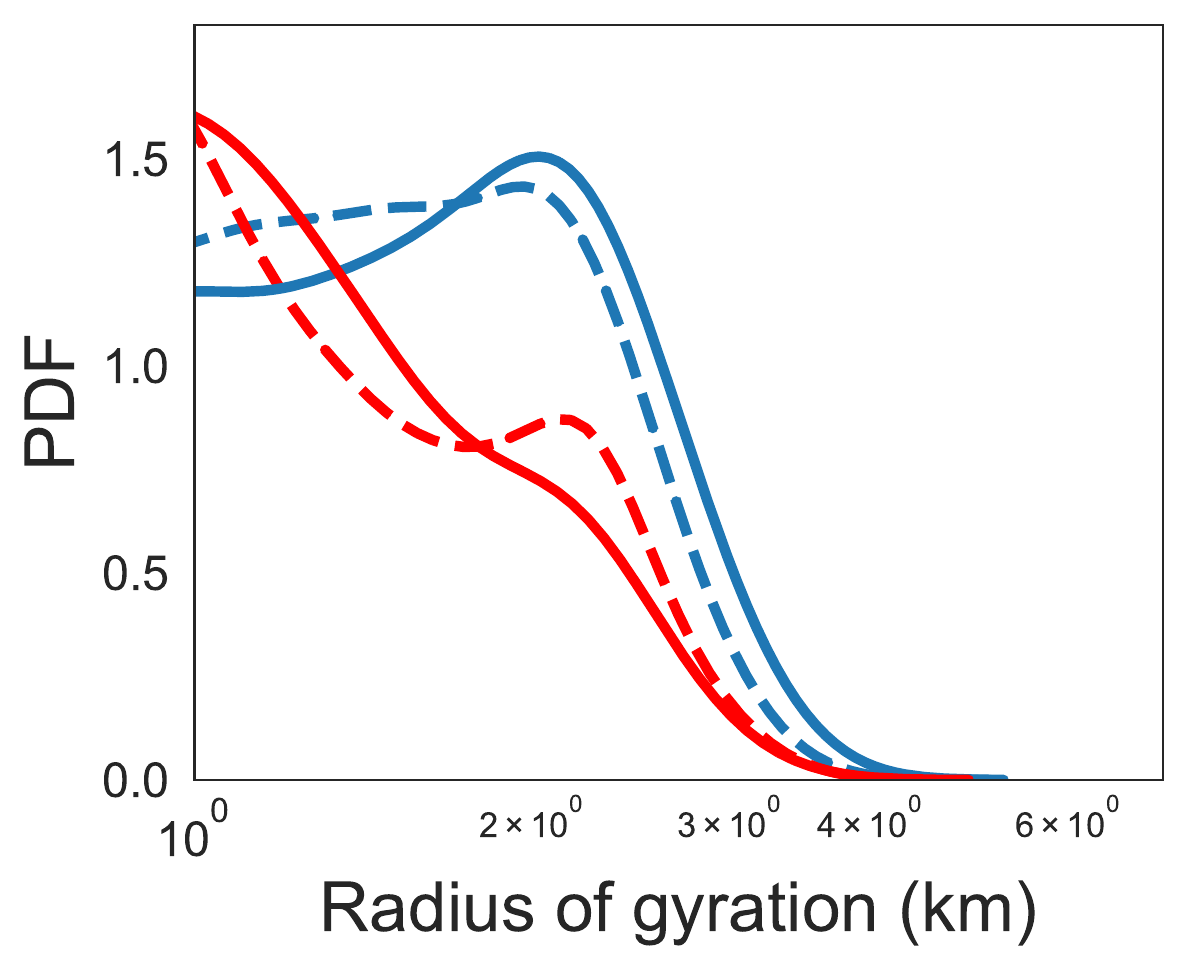}
\caption{} 
\label{fig:radius-gyration-total-ccdf}
\end{subfigure}
\begin{subfigure}{0.19\textwidth}
\includegraphics[scale=0.28]{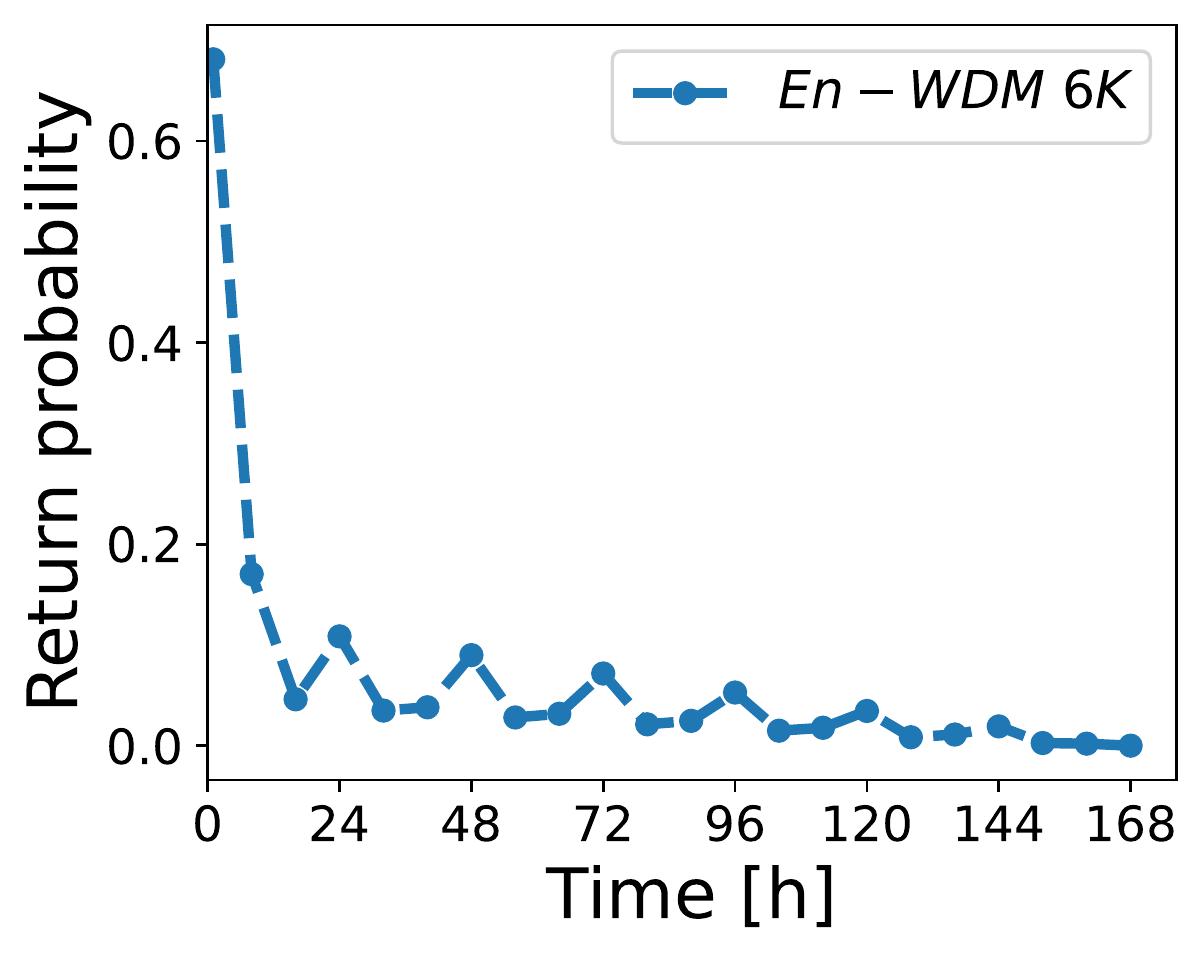}
\caption{} 
\label{fig:return_prob}
\end{subfigure}
\begin{subfigure}{0.19\textwidth}
\includegraphics[scale=0.28]{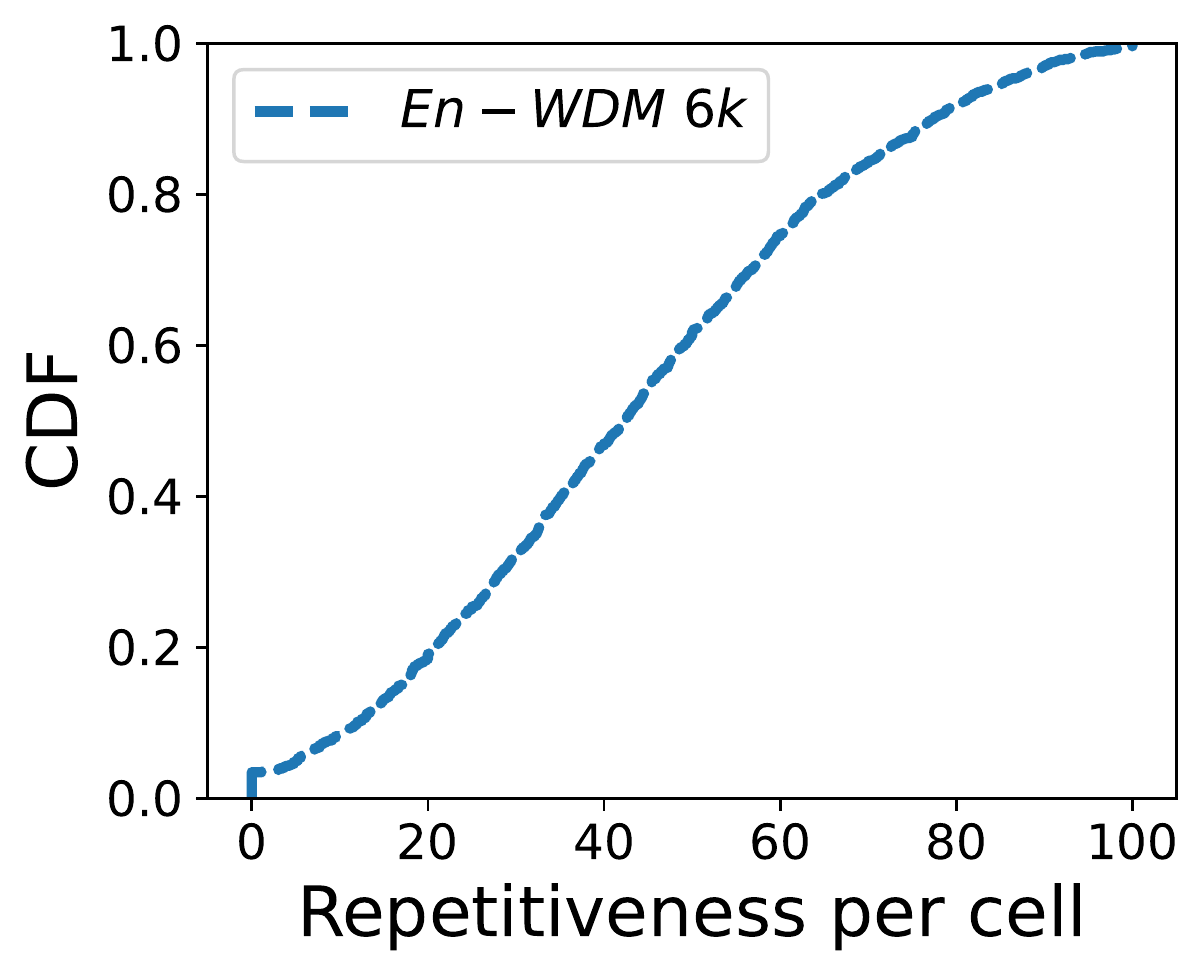}
\caption{} 
\label{fig:repetitiveness_cdf}
\end{subfigure}
\caption{Mobility metrics compared with initial WDM: (a) Inter-contact time CCDF (b) Normalized number of contacts per hour (c) Radius of gyration CDF (d) Return probability (e) Per-cell repetitiveness CDF}
\label{fig:mobility-comparison}
\end{figure*}

Fig. \ref{fig:mobility-comparison} shows well-known metrics for  characterization of wireless networking meetings (inter-contact and contact time) and the tendencies in human mobility, i.e., confinement (radius of gyration) and repetitiveness (probability to return to previously visited places). As for WDM, we emulate a scenario with 1000 and 6000 users, moving in the Helsinki city center area with roughly 7×8.5km for $5.10^5s$ and with the same arrangement of home/work and POIs. 
We use the same representation of results for comparison reasons.

We can see that \textit{En-WDM}'s inter-contact time distribution (cf. Fig. \ref{fig:inter-contact-ccdf}) and the normalized number of contacts (Fig. \ref{fig:contact-per-hour-ccdf}) closely follows the ones of WDM, attesting the realistic modeling of such metric at population scale and the capability of reproducing heterogeneity to mobility decisions.\\
At last, we evaluate the capability of the two models in reproducing seminal literature analytical human mobility laws~\cite{Marta:2008,Amichi:2020,Eduardo_Mob:2016}. The radius of gyration (Fig. \ref{fig:radius-gyration-total-ccdf}) estimates 
the area size mostly covered by daily displacements of a user. In \textit{En-WDM}, the radius of gyration is globally smaller due to routiners and regulars (79.73\% of the population) who have more confined displacements, consistent with real-life mobility behavior~\cite{Amichi:2020}. Moreover, the average return probability (Fig.~\ref{fig:return_prob}) and per-cell repetitiveness ((Fig.~\ref{fig:repetitiveness_cdf}) results show that users have a regular and periodical spatial mobility behavior with a higher probability of returning to a previous small set of visited locations, as shown in~\cite{Marta:2008, Eduardo_Mob:2016}. 

\subsection{\textit{Zen} CdRs use cases}
We evaluate the complete CdRs resulting from \textit{Zen} framework as compared to \textit{RefCdRs} when applied to three use cases. 
As \textit{RefCdRs} lack ground-truth in mobility information, we enrich them with \textit{Zen} CdRs' emulated user trajectories using \textit{Zen}'s CdR-combiner methodology (ref. \S \ref{sec:zen-merger}), we name it \textit{M-RefCdRs}. Based on the confirmed \textit{Zen} performance in reproducing human mobility laws, we focus our use-cases analysis on the reproduction of cellular traffic behavior for which we have a ground-truth.
We generate \textit{Zen} CdRs with 6000 users, corresponding to the same number of users in \textit{RefCdRs} (see \S \ref{subsubsec:datasets}) and consider a week-long period. 

\begin{figure}
\includegraphics[scale=0.12]{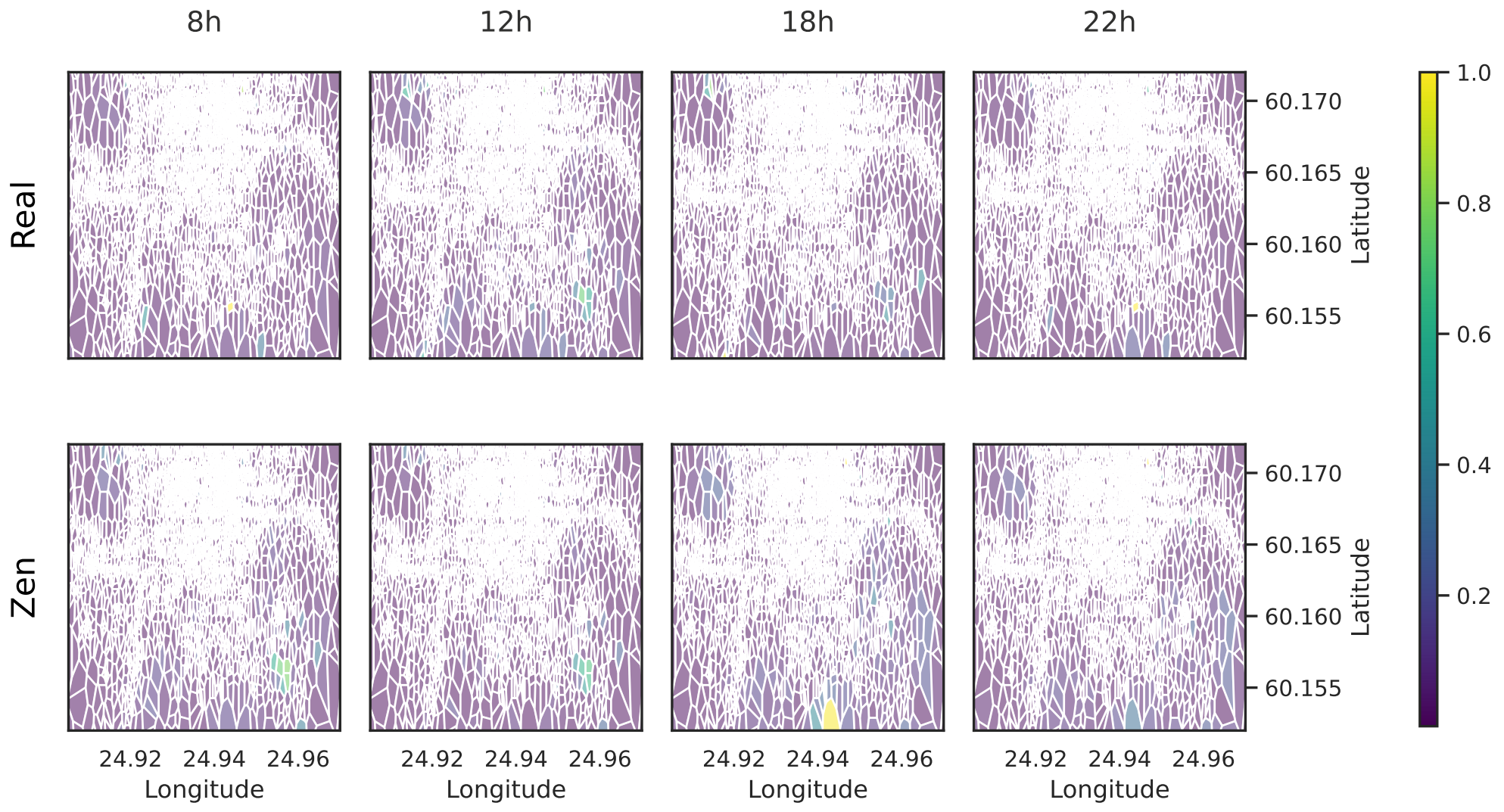}
.\caption{Dynamic people presence estimated at four daily time in Helsinki for real-world and \textit{Zen} traffic.}
\label{fig:voronoi_hour}
\end{figure}

\vspace{0.1cm}
\noindent\textbf{\textit{Dynamic urban tracking.}}
Real-time population density tracking is a key functionality to support adaptive urban and transport planning.
As shown in \cite{Khodabandelou:2019}, such density at time $t$ can be derived from the corresponding network activity load at $t$ computed as the mean number of network events (e.g., here ongoing calls, exchanged SMS, and established data sessions) per individual.
Following this methodology, Fig. \ref{fig:voronoi_hour} shows the spatial distribution (values in the color bar) of people presence in network cells of an Helsinki area (2.2km $\times$ 3.6km), at four representative time hours of individuals' routine, obtained with \textit{M-RefCdRs} and \textit{Zen} CdRs. 
As in \textit{M-RefCdRs}, we see that people presence at the office period (8h-12h) is concentrated in specific zones corresponding to defined Helsinki business neighborhoods. In contrast, the after-work period (18h-22h) includes displacements times and night activities not made at specific spots (e.g., groups of users can walk down the streets for their night activity), explaining people presence is spread over a broader zone. 
Besides, we notice that people presence is captured equivalently in \textit{M-RefCdRs} and \textit{Zen}'s CdRs, especially in working period (8h-12h). We believe the resulting few dissimilarities, particularly for the after-work period (18h-22h), are mainly due to the non-deterministic association of user's traffic to trajectories in Zen's CdR-combiner.

\begin{figure}
\hspace*{-0.7cm}
\begin{subfigure}{0.21\textwidth}
\includegraphics[scale=0.2]{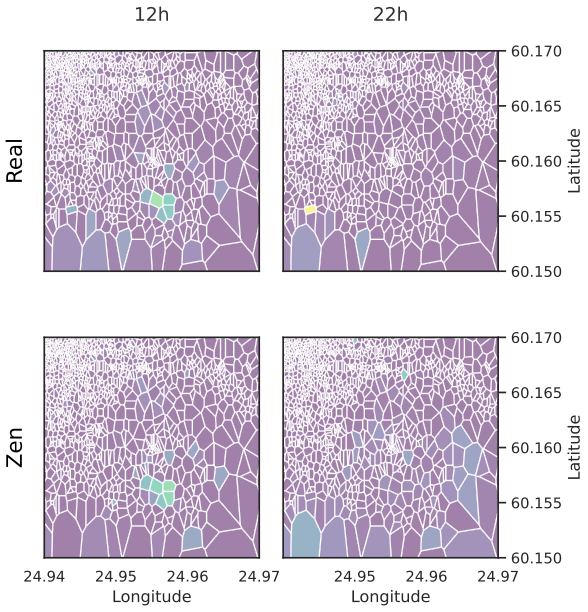}
\caption{Always-active micro BS.} 
\label{fig:}
\end{subfigure}
\begin{subfigure}{0.21\textwidth}
\includegraphics[scale=0.2]{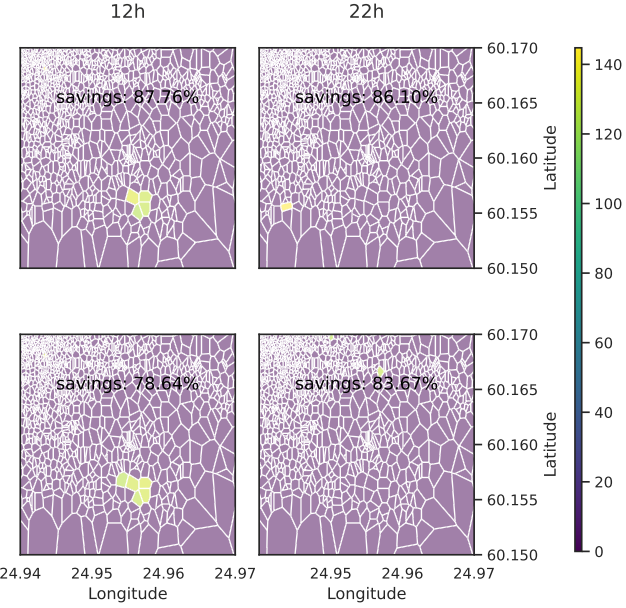}
\caption{Cell-sleeping strategy.} 
\label{fig:}
\end{subfigure}
\caption{Power consumption per cell (a) for always-active micro BS and (b)
with a cell-sleeping strategy. }
\label{fig:power_consumption}
\end{figure}

\vspace{0.1cm}
\noindent\textbf{\textit{Data-Driven Micro BS Sleeping.}}
Numerous works studied power savings in Radio Access Networks (RAN). Inspired by \cite{SpectraGAN}, we investigate how a traffic-aware Base Station (BS) on/off-switching strategy~\cite{Vallero:2019} performs when informed with \textit{Zen} CdRs compared to \textit{M-RefCdRs}. 
We assume an heterogeneous RAN deployment where each cell is served by a separate micro BS, whereas macro BSs provide umbrella coverage to a larger area. Specifically, we consider a grid tessellation of 5X5 macro BSs in the considered zone. The power needed to the operation of a BS at time $t$ is $P(t) = N_{trx} (P_0 + \Delta_p P_{max} \rho (t) )$, $0 \leq \rho (t) \leq 1$, where $\rho (t)$ is the relative traffic load at time $t$ with $P_0, N_{trx}, P_{max}$ and $\Delta_p$ being constants defined for micro and macro BSs in \cite{SpectraGAN}.  Then, if $\rho(t) \leq \rho_{min} = 0.37$ as considered in \cite{Dalmasso:2016} the micro BS offloads its local traffic to the macro BS and goes into sleep
mode, where it consumes negligible power.
Accordingly, Fig. \ref{fig:power_consumption} shows the power consumption ($P(t)$ values in the color bar) of each cell's micro BS at two hours in Helsinki (a zoomed-in area of 2.2km $\times$ 1.6km) with and without such a strategy implemented.  We can see that comparable cells are kept on, while the strategy brings similar energy savings.

\vspace{0.1cm}
\noindent\textbf{\textit{Anomaly detection}.}
Beyond global population-related applications, the fine-grained state of \textit{Zen} CdRs allows for the investigation of per-user spatiotemporal behavior for cellular anomaly detection. 
Such anomalies can be unusual events possibly generated by some security incidents (e.g., stolen account, malware device infection) \cite{Giura:2014} or users with a fraudulent behavior profile. 
As an instance of the latter, SIMBox fraud is a prevalent scam in telecommunication networks consisting of "fake" user accounts re-injecting diverted international calls as local calls to a country \cite{Kouam:2021}. 
We assess the utility of \textit{Zen} CdRs for investigating such fraud by applying a user profiling method where traffic or mobility users' behaviors are leveraged to classify a user as fraudulent or not. 
To this end, we apply for both \textit{Zen} and real ones, a DBSCAN clustering to a set of per-user traffic-related features specific to detect SIMBox fraudulent behavior as described in Table 1 of \cite{Sallehuddin:2015}. Results show a similarity between \textit{Zen} CdRs and real-world ones: while \textit{M-RefCdRs}' estimated number of clusters and outliers are $10$ and $1241$, Zen CdRs' confidence intervals for these metrics are $9.1 \pm 1.66$ and $1122.3 \pm 35.02$ for $10$ samples of \textit{Zen} CdRs' call duration feature (ref. \S \ref{subsec:metric}).



\section{Related literature}
\label{sec:related_works}
CdRs' inaccessibility has pushed researchers to generate their own synthetically, commonly using features modeling. This leads to either mobility- or traffic-specific CdRs, often with grouped-based analysis of individuals' behavior.  \textit{Zen} tackles such lacks by empowering the scientific community with the autonomy needed for the generation of realistic, complete, precise, and flexible CdRs. 

\vspace{0.1cm}
\noindent\textbf{\textit{Traffic-related}}:
Instead of aggregated network traffic generation as done in \cite{Lin:2020, SpectraGAN, CartaGenie}, we focus here on per-individual CdRs generation that tackles different challenges. In this domain, literature's synthetic CdRs lack completeness in describing both call~\cite{Murtic:2018, Milita:2021, Hughes:2019} and mobile data~\cite{Eduardo:2014} usages, and to the best of our knowledge, pay no attention to SMS usage.
Murtić et al.~\cite{Murtic:2018} used Social Network Analysis to reproduce call behaviors' features (i.e., temporal likelihood of calls and call duration distribution) per user profile, extracted from real-world CdRs. Nevertheless, the work did not include any validation. 
In the same vein, Songailaitė et al.~\cite{Milita:2021} statistically model key parameters from real CdRs to produce realistic CdRs. Calling behaviors is simulated based on the empirical fitting of call duration, call count, call likelihood per hour, and weekdays similarity in behaviors. However, simulation relies on a simplistic and randomly-built network social structure leveraging static parameters such as the maximum number of friends and acquaintances. 
Using a GAN generative model, Hughes et al.~\cite{Hughes:2019} show the deep learning models' capability to learn inherent and complex distributions from real CdRs. Unfortunately, real and generated CdRs included only two features: the starting call hour and duration in minutes, revealing a limited extent of modeled features compared to complete CdRs. Finally, Oliveira et al.~\cite{Eduardo:2014} focused on the data-traffic profiling, modeling, and generation from real CdRs. Their model allows generating data usage's timestamped records per profiled user. Although providing flexible settings for profiles' granularity, this work also has the drawback of modeling only data traffic features, lacking thus real-world CdRs' completeness.

\vspace{0.05cm}
\noindent\textbf{\textit{Mobility-related:}}
Synthetically generated mobility traces are regular in literature and frequently extracted from models implemented in ONE~\cite{ONE}, BonnMotion~\cite{BonnMotion}, or SUMO~\cite{SUMO2018} realistic simulators. Several works on mobility modeling actually focus on the generation of synthetic traces that capture specific features in human mobility that are often domain-specific: e.g., MANETS and DTNs (e.g., inter-contact and contact time)~\cite{map-slaw, smooth}, Disaster Management~\cite{Papageorgiou:2012, Aschenbruck:2007} or Sociology~\cite{Borrel:2009}. Still, a few works such as \cite{WDM, RLMM, Pappalardo:2016, Jiang:2016} aim to model real-life mobility and propose more complex models, valuable for more applications. This paper leverages the~\cite{WDM}'s originality in combining various mobility aspects and realistically modeling them. Other strategies rely on recurrent neural networks~\cite{Kulkarni:2017} or statistical generative models based on real mobility traces such as Markov models~\cite{Gonzalo:2017}, spatiotemporal empirical distributions~\cite{WHERE} or travel demand~\cite{Zilske:2014}. Yet, only a few works~\cite{DP-WHERE, Glove} address the privacy issues of generated mobility traces, which is however crucial.

\vspace{-0.1cm}
\section{Conclusion and Discussion}
\label{sec:conclude}
This paper presented \textit{Zen}, the first framework allowing the autonomous generation of complete and realistic CdRs in an individual basis.
To this end, we relied on a fully anonymized and incomplete (only traffic-related) CdRs datasets and provide the first literature modeling that captures long-range and inter-CdRs traffic features correlation, individuals heterogeneity and social-ties in communication. 
The disjoint modeling of realistic emulated mobility and captured real-world traffic behaviors hides real individuals' daily-life habits in routine and leisure times (e.g., home/work, nightlife, etc.), bringing the privacy-preserving capability to the produced \textit{Zen} CdRs. Finally, we validate \textit{Zen} Cdrs (i) realisticness in reproducing daily cellular behaviors of urban population and (ii) usefulness in practical networking applications such as dynamic population tracing, Radio Access Network's power savings, and anomaly detection as compared to real-world Cdrs. Next, we provide extra discussions on possible alternatives and improvements.

\vspace{0.1cm}
\noindent\textbf{\textit{Flexibility and generalization:}}
All the contextual building blocks feeding the \textit{Zen} mobility modeling (e.g., Census information, bus schedule, real city map, neighborhood popularity, etc.) bring generality and flexibility to the representation of city urban life, yielding individuals' cyclic behavior. 
On the other hand, though \textit{Zen} provides realistic traffic behavior models trained from a unique real-world traffic CdRs, the modeling methodology of this paper is general and can be applied to other CdRs with different cultural traffic habits.

\vspace{0.1cm}
\noindent\textbf{\textit{Alternative modeling approaches:}}
LSTMs are perhaps the simplest network (in terms of manual tuning) that can reliably model long-term dependencies and has the flexibility to be used jointly with other more complex architectures. For example, a GAN~\cite{SpectraGAN, GAN_survey_2021} uses paired generator/discriminator networks to enable very realistic output; our work provides the networks that can be used inside the GAN. 

\vspace{0.1cm}
\noindent\textbf{\textit{Future improvements:}}
As mentioned, \textit{Zen} extensively enhances the original WDM model. Nevertheless, as with any research contribution, the mobility generation of \textit{Zen} is still open for improvements, such as the addition of complementary features in fine-grained human mobility laws, cities' contextual information (e.g., friendship, popular leisure zones in the city, etc.), representing weekend mobility and behaviors induced by teleworking or minor users profiles (e.g., unemployed),  or modeling from real-world mobility CdRs.

\vspace{0.1cm}
\noindent\textbf{\textit{Privacy vs individual precision:}}
Although presenting ground- truth modeling and validation opportunities, individual-based mobility modeling of real-world CdRs brings important privacy issues: As users' actual habits in mobility are captured in the model, the generated CdRs have the weakness of revealing aspects in users' daily-life routine, such as important locations (e.g., home/work), regular trajectories (e.g., preferred places for leisure, etc).
The tradeoff between privacy compliance and preciseness in mobility modeling of CdRs is a relevant investigation we let for future work.

\bibliographystyle{ACM-Reference-Format}
\bibliography{reference}

\end{document}